\documentclass[twocolumn]{aastex7}
\usepackage{graphicx}	% Including figure files
\usepackage{amsmath}	% Advanced maths commands
\usepackage{amssymb}	% Extra maths symbols
\usepackage{bm}		% Bold maths symbols, including upright Greek
\usepackage{subfigure}
\usepackage{hyperref}
\usepackage{booktabs}
\usepackage{tablefootnote}
\usepackage{mathtools}
\usepackage{textcomp}

\hypersetup{bookmarksopen=true,bookmarksnumbered=true}

\newcommand{\comment}[1]{}
\newcommand{\PA}{\ensuremath{\Theta}}
\newcommand{\PI}{\ensuremath{\emph{PI}}}
\newcommand{\PF}{\ensuremath{\emph{PF}}}
\newcommand{\PIdeb}{\ensuremath{PI_{deb}}}
\newcommand{\PFdeb}{\ensuremath{PF_{deb}}}
\newcommand{\bpa}{\ensuremath{\theta}}
\newcommand{\NHtwo}{\ensuremath{\mathrm{N_{{\scriptscriptstyle \mathrm{H2}}}}}}
\newcommand{\nHtwo}{\ensuremath{\mathrm{n_{{\scriptscriptstyle \mathrm{H2}}}}}}
\newcommand{\msun}{\ensuremath{\mathrm{M_\odot}}}
\newcommand{\kmps}{\ensuremath{\mathrm{km\,s^{-\!1}}}}
\renewcommand{\micron}{\ensuremath{\,\text{\textmu m}}}
\newcommand{\Bpos}{\ensuremath{\mathrm{B_{pos}}}}
\newcommand{\BposDCF}{\ensuremath{\mathrm{B}_\mathrm{pos}^\textsc{dcf}}}
\newcommand{\BposST}{\ensuremath{\mathrm{B}_\mathrm{pos}^\textsc{st}}}
\newcommand{\muHtwo}{\ensuremath{\text{\textmu}_{\mathrm{H2}}}}
\newcommand{\sigmaQU}{\ensuremath{\sigma_{\!_\mathrm{QU}}}} %{\sigma_{\textsc{qu}}}
\newcommand{\sigmaQi}{\ensuremath{\sigma_{\!_\mathrm{Q,i}}}}
\newcommand{\sigmaUi}{\ensuremath{\sigma_{\!_\mathrm{U,i}}}}

\shorttitle{Polarisation Study of filament in North Orion B}
\shortauthors{K K Mallick et al.}

\graphicspath{{./}{figures/}}
\begin{document}

\title{Fragmenting Filaments and Evolving Cores - Insights from Dust
       Polarisation Study \\ of a filament in Northern Orion\,B}

%%%%%%%%%%%%%%%%%%%%%%%%%%%%%%%%%%%%%%%%%%%%%%%%%%%%%%%%%%%%%%%%%%%%%%%%%%%%%%%%

\author[orcid=0000-0002-3873-6449]{Kshitiz K. Mallick}
\affiliation{National Astronomical Observatory of Japan, Osawa 2-21-1, Mitaka, Tokyo 181-8588, Japan}
\email[show]{kshitiz.mallick@nao.ac.jp}

\author[0000-0002-1959-7201]{Doris Arzoumanian}
\affiliation{Institute for Advanced Study, Kyushu University, Japan}
\affiliation{Department of Earth and Planetary Sciences, Faculty of Science, Kyushu University, Nishi-ku, Fukuoka 819-0395, Japan}
\affiliation{National Astronomical Observatory of Japan, Osawa 2-21-1, Mitaka, Tokyo 181-8588, Japan}
\email{}

\author[0000-0002-7287-4343]{Satoko Takahashi}
\affiliation{National Astronomical Observatory of Japan, Osawa 2-21-1, Mitaka, Tokyo 181-8588, Japan}
\affiliation{Astronomical Science Program, The Graduate University for Advanced Studies, SOKENDAI, 2-21-1 Osawa, Mitaka, Tokyo 181-8588, Japan}
\email{}

\author[0000-0003-0646-8782]{Ray S. Furuya}
\affiliation{Institute of Liberal Arts and Sciences Tokushima University, Minami Jousanajima-machi 1-1, Tokushima 770-8502, Japan}
\affiliation{National Astronomical Observatory of Japan, Osawa 2-21-1, Mitaka, Tokyo 181-8588, Japan}
\email{}

\author[0000-0002-4093-6925]{Yoshiaki Misugi}
\affiliation{Faculty of Science and Engineering, Kyushu Sangyo University, 2-3-1 Matsukadai, Fukuoka 813-8503, Japan}
\affiliation{National Astronomical Observatory of Japan, Osawa 2-21-1, Mitaka, Tokyo 181-8588, Japan}
\email{}

\author[0000-0001-9368-3143]{Yoshito Shimajiri}
\affiliation{Kyushu Kyoritsu University Jiyugaoka 1-8, Yahatanishi-ku Kitakyushu, Fukuoka, 807-08585, Japan}
\email{}

\author[0000-0002-8557-3582]{Kate Pattle}
\affiliation{Department of Physics and Astronomy, University College London, Gower Street, London WC1E 6BT, UK}
\email{}

\author[0000-0003-4366-6518]{Shu-ichiro Inutsuka}
\affiliation{Department of Physics, Graduate School of Science, Nagoya University, Furo-cho, Chikusa-ku, Nagoya 464-8602, Japan}
\email{}

%%%%%%%%%%%%%%%%%%%%%%%%%%%%%%%%%%%%%%%%%%%%%%%%%%%%%%%%%%%%%%%%%%%%%%%%%%%%%%%%

\begin{abstract}

We present an analysis of polarised dust emission at 850\micron\, for a parsec
long filament in the northern part of the Orion\,B molecular cloud. The region
was observed by the JCMT SCUBA-2/POL-2 polarimeter.
The filament has a line mass ($\sim$80\,\msun\,pc$^{-1}$) larger than the
critical (magnetic) line mass ($\sim$37\,\msun\,pc$^{-1}$); and hosts one
starless, three prestellar, and four protostellar cores, with masses in the
range 0.13 to 9.13\,\msun.
The mean (debiased) polarisation fraction of the filament and core pixels was
calculated to be 5.3$\pm$0.3\% and 3.2$\pm$0.3\%, respectively, likely
reflecting their distinct physical conditions.
The polarisation fraction for the cores does not depend on the type of core,
and was found to decrease with increasing column density, varying from 6-11\%
at the filament edges to 1$^{+0.7}_{-0.1}$\% in the denser parts
(\NHtwo$\gtrsim$2$\times$10$^{22}$cm$^{-2}$).
Magnetic field orientation of the protostellar cores, in contrast to prestellar
cores, appears to be relatively aligned with the magnetic field orientation of
the local filament in this region.
Using the Davis-Chandrasekhar-Fermi formalism the plane-of-sky magnetic field
strength for the protostellar cores ($\sim$39-110\,$\text{\textmu}$G) was
found to be higher than that of the prestellar
cores ($\sim$22-61\,$\text{\textmu}$G); and weakest for the
starless core ($\sim$6\,$\text{\textmu}$G).
The average value for the filament was found to be $\sim$31\,$\text{\textmu}$G.
The magnetic field-volume density relation for the prestellar/starless cores
and protostellar cores suggests a transition from weak field case to strong
field case as the cores evolve from prestellar to protostellar phase.
%Context, Aims, Methods, Results, Conclusions
\end{abstract}

\keywords{Submillimeter astronomy (1647), Interstellar magnetic fields (845),
Interstellar filaments (842), Polarimetry (1278), Star formation (1569)}

%%%%%%%%%%%%%%%%%%%%%%%%%%%%%%%%%%%%%%%%%%%%%%%%%%%%%%%%%%%%%%%%%%%%%%%%%%%%%%%%

\section{Introduction}
\label{section_introduction}

Star formation is a complex phenomena requiring simultaneous examination of
multiple competing processes such as gravity, turbulence, and magnetism
\citep{Hennebelle_2019FrASS,Li_PPVI_2014prpl,Elmegreen_2004ARAA,MaclowKlessen_2004RvMP}.
While gravity and turbulence have been studied to a significant degree
\citep{Hennebelle_2012AAR}, investigation of magnetic field is work in progress
\citep{Pattle_PPVII_2023ASPC}, driven over the past decade by developments
in instrumentation \citep{PattleFissel_2019FrASS},
enabling polarisation measurements to catch up with photometric and
spectroscopic observations.
Observations by \emph{Planck} \citep{Planck_XIX_2015AA} have already shown the
ubiquitous presence of (dust) polarised emission throughout the Galaxy,
indicating the magnetized nature of the interstellar medium (ISM).
The first results from \emph{Planck} demonstrated the close relationship
between the magnetic field structure and the density structures, suggesting
the crucial role of the magnetic fields in regulating the dynamics of the ISM
\citep{Planck_XXXII_2016AA,Planck_XXXIII_2016AA,Planck_XXXIV_Rosette_2016AA,
Planck_XXXV_2016AA}.

Analysis of magnetic field associated with Galactic star formation requires
observations of cold molecular clouds as well as structures present therein,
such as filaments, hubs, ridges \citep{Myers_2009ApJ,Hacar_PPVII_2023ASPC},
with scales ranging from a few parsec to tenths of parsec
\citep{Koch_W51_2022ApJ,Pattle_PPVII_2023ASPC}.
Examination of filaments especially assumes relevance as \emph{Herschel}
far-infrared (FIR) observations of dust emission have revealed the ISM
to be threaded by filamentary structures \citep{Andre_PPVI_2014prpl}.
Filaments have also been hypothesised to be an intermediate step in the
collapse of molecular clouds towards dense cores
\citep{InutsukaMiyama_1997ApJ}, and their importance in star formation is
borne out by the significant number of cores found associated with them
\citep{Andre_HGBS_2010AA, Konyves_2020AA_HGBS, Nanda_2020AA}.
The latter suggests that filament evolution may be significantly
intertwined with the initial conditions of star formation.
Hence, it is imperative to study magnetic fields associated with filaments,
particularly those hosting dense cores, in order to understand their dynamical
evolution and fragmentation into cores.

In this paper, we have carried out a polarisation study of a relatively
secluded filament
%\object{OriBupfil}
\citep{Arzoumanian_OriBupfil_2017MmSAI},
unassociated with the main molecular complex, in the northern part of the
\object{Orion\,B} molecular cloud complex.
The location is associated with the \emph{Planck} Galactic Cold Clump
(PGCC) \object{G203.21-11.20}
\citep{Planck_XXVIII_PGCC_2016AA,Planck_VIIEarlyRelease_2011AA},
and has only been a part of a few statistical studies of Orion at FIR and
submillimeter wavelengths. Observations by JCMT (James Clerk Maxwell Telescope)
at 850\micron\, \citep{Yi_2018ApJS}
found four cores associated with the PGCC, named as
\object{G203.21-11.20E1},
\object{G203.21-11.20E2},
\object{G203.21-11.20W1}, and
\object{G203.21-11.20W2}.
According to ALMA observations, G203.21-11.20W2 has been classified as a hot
corino \citep{Liu_ALMASOP_2025ApJ,Hsu_ALMASOP_2022ApJ}.
Dense gas tracers such as N$_2$H$^+$, HCO$^+$, H$^{13}$CO$^+$, and HCN
\citep{Yi_LambdaOrionisIII_2021ApJS,Yuan_2016ApJ}, and isotopologues of CO
\citep{Liu_PCC_Orion_2012ApJS} have also been detected at the location of
\object{PGCC G203.21-11.20}.
Despite the above, while the Orion\,B cloud complex is a well-investigated
high-mass star forming region \citep{Bally_Orion_2008hsf1}, this particular
filament shows a lack of investigation in literature, lacking even
observations in \emph{Spitzer} \citep{Werner_Spitzer_2004ApJS}.
More recently, the filament was found to harbour multiple cores in the
Herschel Gould Belt Survey
\citep[HGBS,][]{Andre_HGBS_2010AA,Konyves_2020AA_HGBS}.
The core masses from the HGBS catalog
(for the filamentary region studied in this paper)
cover nearly two orders of magnitude, ranging from 0.13-9.13\,\msun.
Thus this filament presents a suitable location to
study the magnetic field structure across a filament fragmenting into
cores at different evolutionary stages. 
The above mentioned molecular
tracer studies show the systemic velocity of the PGCC associated with the
filament to be $\sim$10\,\kmps, similar to the rest of the Orion\,B complex
\citep{Pety_OrB_2017AA,Gaudel_OrB_2023AA},
and thus we adopt a distance of 400\,pc usually used for Orion\,B and also
used in the HGBS study \citep{Konyves_2020AA_HGBS}, for our analyses.

We present the 850\micron\, dust polarisation observation carried out
using POL-2 polarimeter on Submillimeter Common-User Bolometer Array 2
(SCUBA-2) on the JCMT \citep{Holland_SCUBA2_2013MNRAS}.
The thermal dust emission at FIR is theorised to be linearly polarised
in a direction perpendicular to the direction of magnetic field due to
alignment of dust grains with their major axis perpendicular to the field
\citep{Andersson_GrainAlignment_2015ARAA,Weintraub_PPIV_2000prpl,
WardThompson_BISTRO_2017ApJ}.
The magnetic field properties can be thus inferred from such observations for
analysis.
The organisation of the paper is as follows :
in Section \ref{section_data}, we list the datasets used in this paper;
in Section \ref{section_results_Polarisation}, we present the polarisation and
magnetic field analyses carried out;
in section \ref{section_results_FilamentAndCores}, we analyse the implications
of the results for the filament and the cores in our region;
in section \ref{section_discussion}, we discuss the overall scenario in the
region; and
finally we summarise the main results in Section \ref{section_summaryconclusions}.

%%%%%%%%%%%%%%%%%%%%%%%%%%%%%%%%%%%%%%%%%%%%%%%%%%%%%%%%%%%%%%%%%%%%%%%%%%%%%%%%

\begin{figure*}[ht]
\centering
%\hspace{-25pt}
\includegraphics[width=0.8\textwidth]{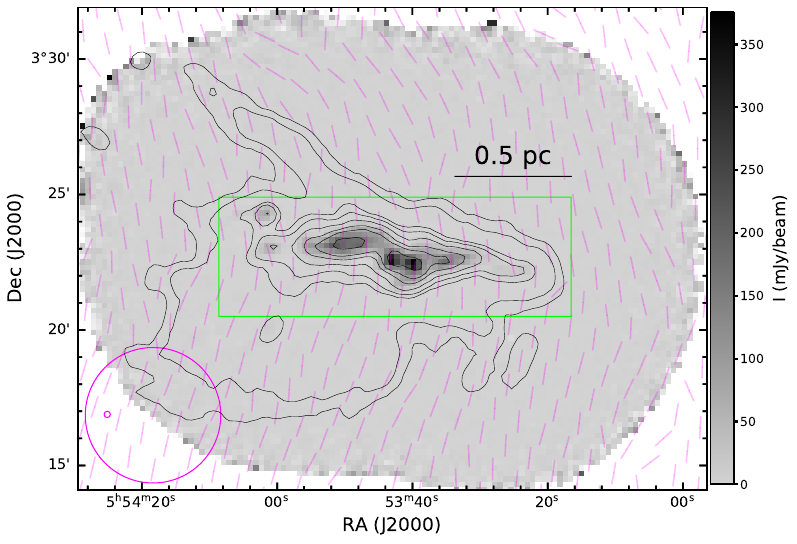}
\caption{
JCMT stokes I map of the observed region. Contours from the HGBS column density
map (resolution$\sim$18.2\arcsec) are drawn at
0.2, 0.25, 0.55, 0.7, 1, 1.5, 2, and 3$\times$10$^{22}$\,cm$^{-2}$.
\emph{Planck} magnetic field half-vectors are shown by magenta line segments.
It should be noted that the \emph{Planck} data is oversampled here.
Magenta circles on bottom left show the \emph{Planck} beam ($\sim$5\arcmin)
and the JCMT beam ($\sim$14\arcsec).
Green box is the filamentary region analysed in this work.
}
\label{fig_StokesI_FullFoV}
\end{figure*}

\section{Data}
\label{section_data}

%%%%%%%%%%%%%%%%%%%%%%%%%%%%%%%%%%%%%%%%

\subsection{JCMT POL-2 Data}
\label{section_data_Pol2}

Polarised dust emission towards the filament was observed under project
codes M20AP021 and M20BP044 (PI: K. Pattle), using SCUBA-2/POL-2 on the JCMT
\citep{Bastien_POL2_2011ASPC,Holland_SCUBA2_2013MNRAS,Friberg_POL2_2016SPIE}.
The standard SCUBA-2/POL-2 DAISY mapping mode was used to carry out the
observations. Two maps of about 17\arcmin\, in diameter each were taken to
cover the eastern and western parts of the pc-long elongated filamentary
structure. Data reduction was carried out in a manner similar to
\citet{Arzoumanian_NGC6334_2021AA}, using \textsc{starlink}'s
\emph{pol2map}\footnote{https://starlink.eao.hawaii.edu/docs/sc22.htx/sc22.html}
data reduction pipeline
\citep{Chapin_SCUBA2_2013MNRAS,Currie_Starlink_2014ASPC}.
The two subfields were combined in a mosaic map of
size $\sim$23\arcmin$\times$18\arcmin.
The JCMT has a diameter of 15\,m, enabling it to achieve an effective
half-power-beam-width (HPBW) resolution of 14\arcsec\, at 850\micron.
The data was projected onto a 12\arcsec\, pixel grid for further analysis.

Figure \ref{fig_StokesI_FullFoV} shows the stokes I image of the entire
observed FoV (field-of-view) by JCMT. Column density contours from HGBS
delineate the morphology of the cloud complex. The filamentary region
enclosed in the green box was analysed in this work.
The mean standard deviation of the stokes I, Q, and U images (for the
analysed region) was found to be
$\sim$0.66, 0.71, and 0.7 mJy\,beam$^{-1}$, respectively.

%%%%%%%%%%%%%%%%%%%%%%%%%%%%%%%%%%%%%%%%

\subsection{Archival Data}
\label{section_data_archival}

Apart from the JCMT POL-2 data, we also utilise \emph{Herschel} column density
(\NHtwo) map from \citet{Konyves_2020AA_HGBS} and polarisation stokes images
from \emph{Planck} \citep{Planck_I_2016AA}.

\citet{Konyves_2020AA_HGBS} have utilised the FIR data in 160-500\micron\,
range from
HGBS\footnote{http://gouldbelt-herschel.cea.fr/archives}
to obtain
\textquotedblleft high-resolution\textquotedblright\, (18.2\arcsec)
column density map of the Orion\,B molecular cloud complex.
We retrieved the map from the VizieR
archive\footnote{https://cdsarc.cds.unistra.fr/ftp/J/A+A/635/A34/}.
The HGBS image was reprojected and cropped to the pixel grid and FoV of our
stokes I image.
\citet{Konyves_2020AA_HGBS} also provide a catalog of cores, classified into
\textquotedblleft starless\textquotedblright,
\textquotedblleft prestellar\textquotedblright, and
\textquotedblleft protostellar\textquotedblright\, categories. Briefly, the
detected dense cores have been classified as protostellar in the catalog if
they have detections at 70\micron, but as starless/prestellar cores if they
do not. Sub-classification between starless and prestellar cores is based on
the analyses differentiating if they are self-gravitating or not. While
multiple criteria have been employed by \citet{Konyves_2020AA_HGBS} in their
catalog, for the starless/prestellar sources in our FoV,
the distinction is based on the ratio of thermal Bonnor-Ebert critical mass
to the core mass derived from \emph{Herschel} observations.
Cores (i.e. those with non-detection at 70\micron) with this ratio $\leq2$
were taken to be self-gravitating. As self-gravitating cores could possibly
evolve into protostars, they have been marked as
\textquotedblleft prestellar\textquotedblright\, in the catalog.
According to ALMA observations
\citep{Dutta_ALMASOP_2020ApJS, Jhan_ALMASOP_2022ApJ, Liu_ALMASOP_2025ApJ},
two sources,
(G203.21-11.20W1 and G203.21-11.20W2 corresponding to cores 5 and 6,
respectively, in this work; see Section \ref{section_polarisationMorphology}),
which have been classified as
\textquoteleft prestellar\textquoteright\, and
\textquoteleft protostellar\textquoteright\,
in the HGBS catalog were found to be associated with outflows.
Since outflows
are a feature from protostellar stage onwards, we changed the designation of
the associated \textquoteleft prestellar\textquoteright\, source from HGBS to
\textquoteleft protostellar\textquoteright.

Besides the above datasets, \emph{Planck} cutouts (of stokes Q and U)
for the region at 353\,GHz were obtained from the IRSA
archive\footnote{https://irsa.ipac.caltech.edu/Missions/planck.html}
\citep{Planck_I_2016AA}. The data was mainly used for a visual analysis
of the large scale orientation of the magnetic field
(see Appendix \ref{appendix_Planck} for calculation details).

%%%%%%%%%%%%%%%%%%%%%%%%%%%%%%%%%%%%%%%%%%%%%%%%%%%%%%%%%%%%%%%%%%%%%%%%%%%%%%%%

\section{Polarisation Results}
\label{section_results_Polarisation}

%%%%%%%%%%%%%%%%%%%%%%%%%%%%%%%%%%%%%%%%

\subsection{Calculation of Polarisation Parameters}
\label{section_PolParams}

Using the stokes I, Q, and U data, and their respective variances
(Section \ref{section_data_Pol2}), the observed polarisation
parameters -- polarisation intensity (\PI), polarisation fraction (\PF), and
the polarisation angle (\PA) are calculated using the following expressions :
\begin{eqnarray}
&\PI =& \sqrt{Q^{2} + U^{2}} \\
&\PF =& \frac{\PI}{I} \\
&\PA =& \mathrm{\frac{1}{2}~tan^{-1}}\frac{U}{Q} \label{eqn_pa}
\end{eqnarray}
While U and Q are scalars, for \PA\, to match with the sign convention of
the polarised waves
\citep{Chandrasekhar_1950ratr,Hamaker_StokesIAU_1996AAS, Wilson_2013tra},
the quantity (U/Q) is not treated as a pure ratio, but the signs are taken
into account to map the angle in the
\textquotedblleft correct quadrant\textquotedblright.
The polarisation angles lie in the range (-90,90) degrees. It
should be noted that according to the IAU convention, angles are measured east
of north. In the FIR regime, the orientation of the plane-of-sky magnetic field
(\bpa) is subsequently obtained by rotating \PA\, by 90\textdegree\,
\citep{Planck_XIX_2015AA, Hildebrand_2000PASP}, i.e.
\begin{equation}
\label{eqn_bpa}
\bpa = \PA + \pi/2.
\end{equation}
To account for the positive bias of the polarisation measurements
\citep{Vaillancourt_2006PASP}, we calculated the debiased polarisation
intensity and polarisation fraction as follows :
\begin{eqnarray}
\PIdeb &=& \sqrt{Q^{2} + U^{2} - 0.5\,(\sigma_{_Q}^2 + \sigma_{_U}^2)}, \\
\PFdeb &=& \frac{\PIdeb}{I}.
\end{eqnarray}
Error propagation yields the following expressions for the uncertainties :
\begin{equation}
\label{eqn_ErrorBpa}
\sigma_{\scriptscriptstyle \bpa} = \sigma_{\scriptscriptstyle \PA} = \frac{1}{2} \frac{\sqrt{Q^2 \sigma_{_U}^2 + U^2 \sigma_{_Q}^2}}{Q^2 + U^2}
\end{equation}
\begin{equation}
\sigma_{_{\PIdeb}} = \frac{\sqrt{Q^2 \sigma_{_Q}^2 + U^2 \sigma_{_U}^2}}{\PIdeb}
\end{equation}
\begin{equation}
\sigma_{_{\PFdeb}} = \PFdeb \sqrt{\frac{\sigma_{_{\PIdeb}}^2}{\PIdeb^2} + \frac{\sigma_{_I}^2}{I^2}}
\end{equation}

Having obtained a map of \PIdeb, \PFdeb, and \bpa, the subsequent task is to
discern reliable measurements for further analysis. Towards this end, we use
stokes I and \PFdeb. Since pixels for which \PFdeb$>$20\% are likely
unrealistic \citep{Montier_BestEstmII_2015AA,Benoit_2004AA,Planck_XIX_2015AA},
we confine our pool of usable pixels to those with \PFdeb$\leq$20\%.
\PFdeb\, for
the region was found to decline with increasing SNR (signal-to-noise
ratio), and hence we choose a liberal criteria of SNR(\PFdeb)$\geq$1 for
selecting our pixels
(we use the notation SNR(\PFdeb)$\equiv$\PFdeb/$\sigma_{\PFdeb}$).
The validity of this selection criteria is discussed in
Appendix \ref{appendix_compareSNRlimits}.
All such usable pixels were found to have a high SNR in I -- SNR(I)$\geq$9 --
hence no further trimming on the basis of SNR of stokes I was deemed necessary.

To summarise, we obtain 198 pixels
(pixel size$\sim$12\arcsec, beam size$\sim$14\arcsec)
in our region of interest which satisfy the combined criteria :
\PFdeb$\leq$20\%, SNR(I)$\geq$9, and SNR(\PFdeb)$\geq$1.

%%%%%%%%%%%%%%%%%%%%%%%%%%%%%%%%%%%%%%%%

%\comment{
%\begin{rotatetable}
\begin{deluxetable*}{cp{3cm}|r|rrr|rrrr}[ht]
%\tabletypesize{\scriptsize}
\digitalasset
\tablewidth{0pt}
\tablecaption{Polarisation Parameters for the extended-core regions \label{table_extCoreSummary}}
\tablehead{
\colhead{S No.}  & \colhead{Quantity}  &
\colhead{Core 8} & \colhead{Core 1} & \colhead{Core 3} & \colhead{Core 4} &
\colhead{Core 2} & \colhead{Core 5} & \colhead{Core 6} & \colhead{Core 7}
}
\startdata
\hline
\multicolumn{3}{l}{Basic Parameters} \\
\cmidrule(lr){1-2}
  1 &                                           CoreType &        Starless &      Prestellar &      Prestellar &      Prestellar &    Protostellar &    Protostellar &    Protostellar &    Protostellar  \\
  2 &                                RA$_\mathrm{J2000}$ &     05:53:23.52 &     05:54:01.51 &     05:53:51.58 &     05:53:47.48 &     05:54:00.88 &     05:53:42.67 &     05:53:39.49 &     05:53:35.13  \\
  3 &                               Dec$_\mathrm{J2000}$ &     +03:22:12.2 &     +03:24:19.3 &     +03:23:06.5 &     +03:23:08.9 &     +03:23:03.6 &     +03:22:36.3 &     +03:22:25.3 &     +03:22:34.2  \\
  4 &                   $a_{\mathrm{extCore}}$ (\arcsec) &              27 &              18 &              36 &              30 &           20.25 &           17.25 &              18 &           23.25  \\
  5 &                   $b_{\mathrm{extCore}}$ (\arcsec) &            13.5 &            16.5 &           20.25 &           20.25 &            13.5 &           14.25 &            13.5 &           14.25  \\
  6 &                            PA$_{\mathrm{extCore}}$ &              98 &             110 &              99 &              86 &              70 &              20 &              16 &              88  \\
  7 &                        Pixels$_{\mathrm{extCore}}$ &               7 &               7 &              18 &              11 &               7 &               9 &               9 &               6  \\
  8 &                           Pixels$_{\mathrm{ngbr}}$ &               4 &               2 &              33 &              30 &               2 &              25 &              30 &              21  \\
\hline
\multicolumn{3}{l}{Polarisation Parameters} \\
\cmidrule(lr){1-2}
  9 &               $\theta_{\mathrm{extCore}} (degree)$ &       94$\pm$15 &        47$\pm$4 &        15$\pm$6 &       169$\pm$6 &       110$\pm$3 &        39$\pm$3 &        18$\pm$2 &        65$\pm$6  \\
 10 &                  $\theta_{\mathrm{ngbr}} (degree)$ &       29$\pm$29 &       145$\pm$6 &        78$\pm$4 &        84$\pm$3 &       64$\pm$32 &        34$\pm$7 &        35$\pm$3 &        40$\pm$3  \\
 11 &                       PF$_{\mathrm{extCore}}$ (\%) &    10.9$\pm$2.6 &     6.1$\pm$0.9 &     1.7$\pm$0.1 &     0.9$\pm$0.2 &     6.4$\pm$1.2 &     1.0$\pm$0.1 &     1.1$\pm$0.1 &     1.2$\pm$0.2  \\
 12 &                          PF$_{\mathrm{ngbr}}$ (\%) &    10.1$\pm$3.2 &    13.3$\pm$3.2 &     5.6$\pm$0.6 &     4.7$\pm$0.6 &    12.3$\pm$5.5 &     3.1$\pm$0.5 &     2.7$\pm$0.4 &     3.4$\pm$0.5  \\
\hline
\multicolumn{5}{l}{Magnetic Field Strengths} \\
\cmidrule(lr){1-2}
 13 &         $\delta\theta_{\mathrm{extCore}} (degree)$ &              48 &              20 &              45 &              29 &              13 &              23 &              18 &              32  \\
 14 &     $\Delta\!\mathrm{V_{NT,extCore}} (km\,s^{-1})$ &            0.20 &            0.29 &            0.63 &            0.65 &            0.31 &            0.69 &            0.69 &            0.55  \\
 15 & B$_{\mathrm{pos,extCore}}^{\textsc{dcf}}$ ($\mu G$) &     5.9$\pm$4.1 &   22.4$\pm$11.9 &   37.7$\pm$12.7 &   60.8$\pm$22.3 &   39.4$\pm$20.5 &   87.9$\pm$28.4 &  109.8$\pm$36.0 &   40.0$\pm$15.1  \\
 16 & B$_{\mathrm{pos,extCore}}^{\textsc{st}}$ ($\mu G$) &     7.7$\pm$5.0 &    18.9$\pm$9.4 &   47.9$\pm$15.3 &   62.0$\pm$20.2 &   26.8$\pm$12.6 &   80.2$\pm$24.7 &   89.2$\pm$27.6 &   43.0$\pm$14.7  \\
 17 &            $\delta\theta_{\mathrm{ngbr}} (degree)$ &              56 &               - &              40 &              36 &               - &              45 &              23 &              29  \\
 18 &        $\Delta\!\mathrm{V_{NT,ngbr}} (km\,s^{-1})$ &            0.24 &               - &            0.48 &            0.52 &               - &            0.56 &            0.54 &            0.49  \\
 19 &   B$_{\mathrm{pos,ngbr}}^{\textsc{dcf}}$ ($\mu G$) &     6.5$\pm$4.9 &               - &    25.7$\pm$9.4 &   32.6$\pm$11.4 &               - &   30.5$\pm$11.0 &   56.1$\pm$19.7 &   35.8$\pm$13.1  \\
 20 &    B$_{\mathrm{pos,ngbr}}^{\textsc{st}}$ ($\mu G$) &     9.2$\pm$5.5 &               - &   30.7$\pm$10.9 &   37.3$\pm$12.7 &               - &   38.7$\pm$13.0 &   50.5$\pm$17.0 &   36.5$\pm$12.9  \\
\enddata
\tablecomments{\emph{(Row-wise)}
1 - Type of core from the HGBS catalog \citep{Konyves_2020AA_HGBS} and ALMA
outflow observations \citep{Dutta_ALMASOP_2020ApJS}.
2 - Right Ascension of the core (J2000).
3 - Declination of the core (J2000).
4,5 - Semimajor and Semiminor Axis dimensions, respectively, in arcseconds
corresponding to coresize of 1.5\,\textsc{fwhm} (i.e. the
\textquotedblleft extended-core\textquotedblright\, region,
indicated by the subscript
\textquotedblleft \emph{extCore}\textquotedblright).
\textsc{fwhm} was taken from the HGBS catalog.
6 - Position Angle taken from the HGBS catalog.
7 - Number of pixels which belong to the extended-core region.
8 - Number of pixels which belong to the local neighbourhood of the
extended-core region.
9 - Mean (circular) \Bpos\, orientation angle of the extended-core region.
10 - Mean (circular) \Bpos\, orientation angle of the local neighbourhood of
the extended-core region.
11 - Mean debiased polarisation fraction of the extended-core region.
12 - Mean debiased polarisation fraction of the local neighbourhood of the
extended-core region.
13 - Angular dispersion (here, we use the circular standard
deviation of the angles of the constituent pixels).
14 - Non-thermal linewidth estimated using the empirical relation of
\citet[see their Figure 6 - Total Velocity Dispersion vs. Column Density]{Arzoumanian_2013AA}.
The error in all (i.e. $\sigma_{\scriptscriptstyle \Delta V_{NT}}$)
was assumed to be $\sim$0.12\,\kmps\,($\approx$0.05$\times\sqrt{\mathrm{(8 ln2)}}$,
see Appendix \ref{appendix_Errors_Bpos}).
15 - Magnetic field strength calculated using DCF method for the extended-core.
16 - Magnetic field strength calculated using \citet{SkalidisTassis_2021AA}
method for the extended-core.
17,18,19,20 - same as rows 13,14,15,16 but for the local neighbourhood of the
extended-core regions. Core 1 and 2 lack values as their neighbourhoods contain
only 2 pixels each due to which a measure of angular dispersion cannot be
obtained, and thus we skip their calculation.
}
\end{deluxetable*}
%\end{rotatetable}
%}

\begin{figure*}
\centering
%
%Section 3.2 from https://cs.brown.edu/about/system/managed/latex/doc/subfigure.pdf
\renewcommand{\thesubfigure}{(top)}
\subfigure
{
%\centering
\includegraphics[width=1\textwidth]{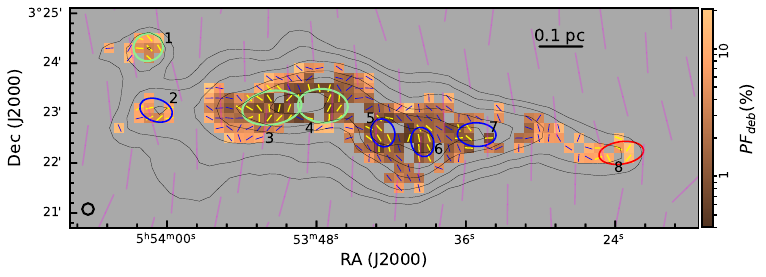}
\label{fig_quiverPlot_PF}
}
\renewcommand{\thesubfigure}{(bottom)}
\subfigure
{
%\centering
\includegraphics[width=1\textwidth]{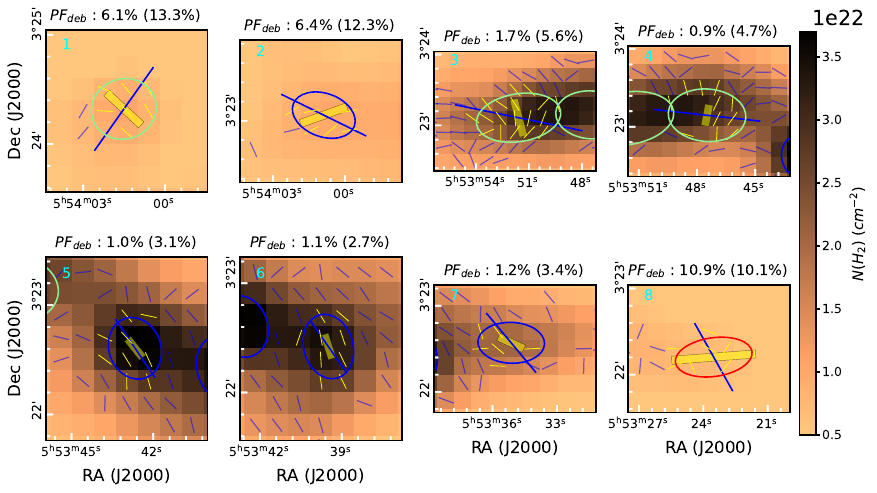}
\label{fig_CoreNbrhoodCompare}
}
\caption{
\emph{(top)}
\PFdeb\, image overlaid with HGBS sources.
Red, Green, and Blue ellipses are the starless, prestellar, protostellar
extended-cores, respectively.
We have labelled these cores 1-8 for ease of analyses. Contours mark the
\NHtwo\, thresholds of 0.55, 0.7, 1, 1.5, 2, and 3$\times$10$^{22}$\,cm$^{-2}$.
Blue and yellow line segments denote the orientation of the POL-2 \Bpos\,
half-vectors (\bpa, measured east of north) for the filament and the cores,
respectively.
Magenta line segments show the magnetic field half-vectors from \emph{Planck}.
It should be noted that the \emph{Planck} data is oversampled here, and
the entire filament is covered by $\sim$3 \emph{Planck} beams ($\sim$5\arcmin).
Black circle on bottom left shows the \emph{JCMT} beam ($\sim$14\arcsec).
\emph{(bottom)} Zoomed-in image of the cores (and their respective
neighbourhoods used for calculation in Section
\ref{section_PolParams_CoreRegions}) overlaid on the column density map.
Thicker blue and yellow segments (in each subfigure) denote the mean \bpa\, of
the neighbourhood (\bpa$_{ngbr}$) and the core (\bpa$_{extCore}$), respectively.
These two segments have been scaled
by \PFdeb; but for cores 1, 2, and 8, the lengths have been further scaled down
by a factor of 3. At the top of each subfigure, mean \PFdeb\, values for the
core and the neighbourhood (in brackets) are given.}
\label{fig_PolMorphologyCompare}
\end{figure*}
%\clearpage

\subsection{Polarisation Morphology}
\label{section_polarisationMorphology}

Figure \ref{fig_quiverPlot_PF} shows the polarisation fraction map of the
region, overlaid with JCMT and \emph{Planck} magnetic field half-vectors
(normalized length), column density contours, and the HGBS cores
(labelled 1-8 from east to west).
To compensate for the sparsity of pixels within the area of
1\,\textsc{fwhm}
coresize \citep[fitted by ][]{Konyves_2020AA_HGBS}, we take
1.5\,\textsc{fwhm}
as the coresize in our analyses. The ellipse sizes
correspond to
1.5\,\textsc{fwhm}
in Figure \ref{fig_quiverPlot_PF}.
To prevent any confusion in terminology related to the area used for different
analyses, we call the region corresponding to 1.5\,\textsc{fwhm} coresize
as \textquotedblleft extended-core\textquotedblright, and affix subscript
\textquotedblleft \emph{extCore}\textquotedblright\, to the parameters derived
using this size.
For the rest of the polarisation parameters' analyses, the
\textquotedblleft extended-core\textquotedblright\, region is used when
analysing the polarisation parameters of the
\textquotedblleft cores\textquotedblright.
It is notable that, in Figure \ref{fig_quiverPlot_PF},
the \emph{Planck} half-vectors appear to be oriented perpendicular to the
longitudinal axis of the filament; and are roughly aligned with the direction
of magnetic field half-vectors of some of the cores (like 3, 4, 5, and 6).
%(also see Section \ref{section_PolParams_CoreRegions}).
The variation of the parameters along and across the filament crest
is discussed in Appendix \ref{appendix_radialVariation}.
The magnetic field orientation of the cores with respect to the filament
crest does not appear to show any particular trend
(see Appendix \ref{appendix_radialVariation_TraversalAlong}).

The area enclosed by each ellipse was visually examined and pixels which seemed
to be largely covered by the ellipses were classified as belonging to the
respective (extended-)core regions, and the rest as belonging to the filament.
Not all pixels belonging to the filament/core regions had detections at our
requisite criteria, and among the total 198 pixels which satisfied the
detection criteria (see Section \ref{section_PolParams}),
125 were classified as filament pixels and 73 as extended-core pixels.

Table \ref{table_extCoreSummary} lists the basic parameters of the
extended-cores (labelled 1-8 from east to west), as well
as the number of detected pixels associated with each extended-core region
(Pixels$_\mathrm{extCore}$) and their respective filamentary neighbourhoods
(Pixels$_\mathrm{ngbr}$, see Section \ref{section_PolParams_CoreRegions}).

%%%%%%%%%%%%%%%%%%%%%%%%%%%%%%%%%%%%%%%%

\begin{figure}
\centering
%
%Section 3.2 from https://cs.brown.edu/about/system/managed/latex/doc/subfigure.pdf
\renewcommand{\thesubfigure}{(a)}
\subfigure
{
%\centering
%\hspace{-35pt}
\includegraphics[width=1\columnwidth]{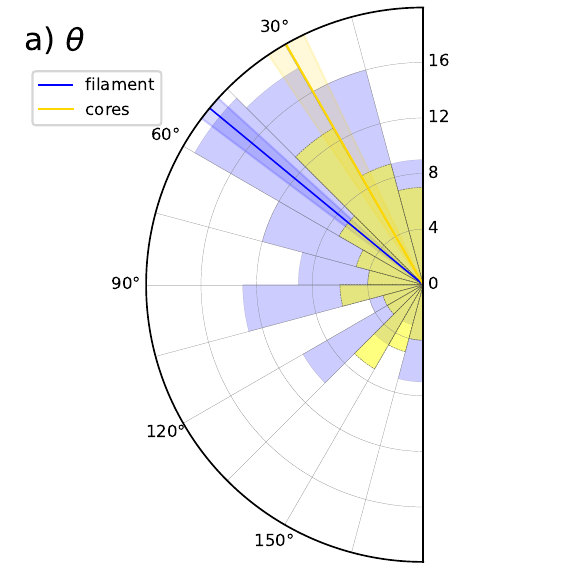}
\label{fig_histogram_byType_Bpa}
}
\renewcommand{\thesubfigure}{(b)}
\subfigure
{
%\centering
\includegraphics[width=1\columnwidth]{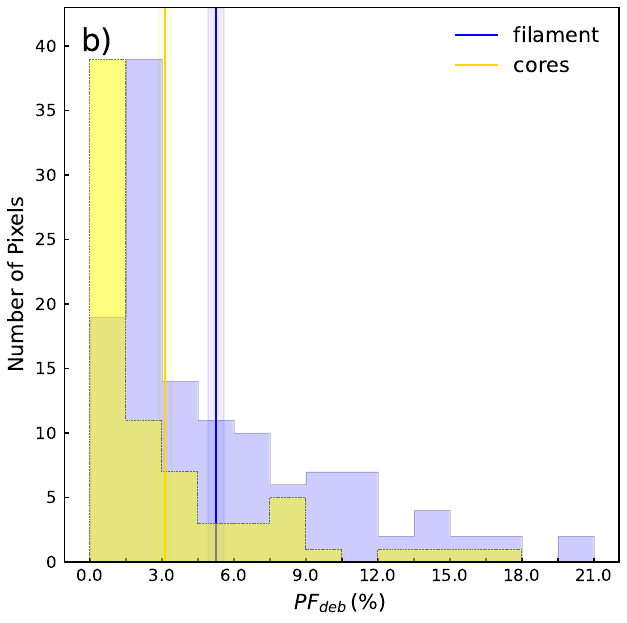}
\label{fig_histogram_byType_pfdeb}
}
\caption{
\emph{(a)} Polar histogram of \bpa\, (orientation angle of \Bpos).
\emph{(b)} Histogram of \PFdeb.
The distributions for
the filament and extended-core regions are shown in
blue and yellow, respectively; and the mean values as
blue lines (50$\pm$3\,degrees and 5.3$\pm$0.3\%) and
yellow lines (30$\pm$4\,degrees and 3.2$\pm$0.3\%), respectively.
The color spans on either side of the mean vertical lines show the error
on the mean.
}
\label{fig_histogram_byType}
\end{figure}
%\clearpage

\begin{figure*}
\centering
%\hspace{-40pt}
\includegraphics[width=0.92\textwidth]{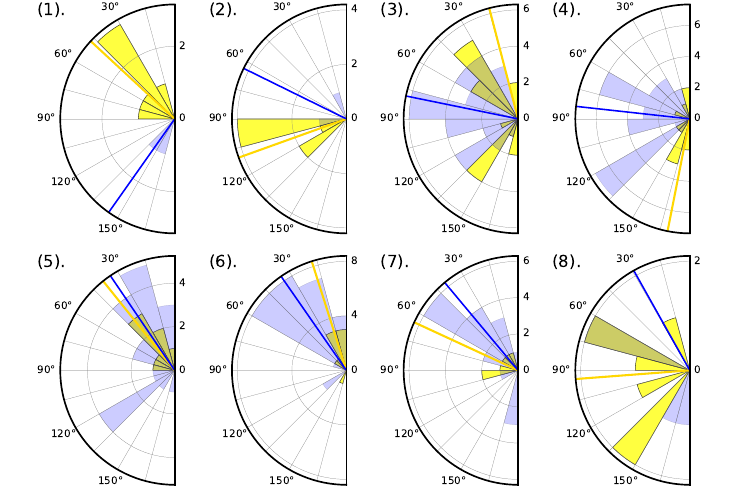}
\caption{
Polar histograms for the eight extended-core regions and respective
neighbourhood regions are shown in yellow and blue, respectively; and
the mean values (see Table \ref{table_extCoreSummary}) as
yellow and blue lines, respectively.
Angles are measured counter-clockwise from 0\textdegree\, at the top.
}
\label{fig_RosePlotCores}
\end{figure*}

\subsection{Histogram of Polarisation Parameters}
\label{section_PolParams_Histogram}

Figure \ref{fig_histogram_byType} shows the distribution of \bpa\, and \PFdeb,
along with their mean values. The data for the filament and the extended-cores
are plotted separately
to investigate whether the properties change from the filament to the cores.
Figure \ref{fig_histogram_byType_Bpa} shows the polar histogram of \bpa\,
%(also called rose plot in literature)
in the 0-180 degree range. As the
magnetic field vectors are half-vectors (also referred to as pseudo-vectors)
-- so called due to a 180\,degree ambiguity and thus an inability to assign
direction --
the same histogram can be shown in any 180\,degree semicircular range.
Since our polarisation angles (\PA) are returned in (-90,90)\,degree range
by Equation \ref{eqn_pa} and \bpa\, is perpendicular to the polarisation
direction (see Equation \ref{eqn_bpa}), we choose (0,180)\,degree range for
consistency.

In the case of \bpa, circular statistics are applicable, and hence the mean
shown is the circular mean. Since \bpa\, data are axial in nature (i.e.
\bpa\, and \bpa+180\textdegree\, are equivalent), they were first converted to
circular (or directional) data by multiplying each angle by two. The circular
mean was then calculated on this dataset of doubled angles, followed by
dividing the so-calculated circular mean by two to obtain the circular mean
of the \bpa\, sample dataset
(see Appendix \ref{appendix_Errors_CircMean} for error calculation).
In Figure \ref{fig_histogram_byType_Bpa}, the circular mean for the filaments
and the cores are shown by blue and yellow lines, respectively.
Figure \ref{fig_histogram_byType_pfdeb} shows the distribution of \PFdeb.
Here, the means shown are the arithmetic means.

The mean of \bpa\, for the filament and the cores is
50$\pm$3\,degrees and 30$\pm$4\,degrees, respectively; while \PFdeb\,
is $\sim$5.3$\pm$0.3\% and $\sim$3.2$\pm$0.3\%, respectively.
The distribution of \bpa\, shows that while the bulk of the angles are
concentrated in the first quadrant, for both the filament and the cores,
the mean direction of the filament is nearly double than that of the cores.
There appears to be a multimodal distribution for both, with other peaks
in the 90-105\,degree and 120-135\,degree sectors.
\PFdeb\, shows a relatively clearer trend, though here also the distribution of
the two are clearly distinct in terms of not only the mean, but also
the peak of the distribution. The means are well-separated by one binwidth
(1.5\%), and thus indicate the distributions to be distinct from each
other.

%%%%%%%%%%%%%%%%%%%%%%%%%%%%%%%%%%%%%%%%

\subsection{Polarisation Parameters of Individual Regions}
\label{section_PolParams_CoreRegions}

\begin{figure}[ht!]
\centering
%
%Section 3.2 from https://cs.brown.edu/about/system/managed/latex/doc/subfigure.pdf
\renewcommand{\thesubfigure}{(a)}
\subfigure
{
%\centering
\includegraphics[width=1\columnwidth]{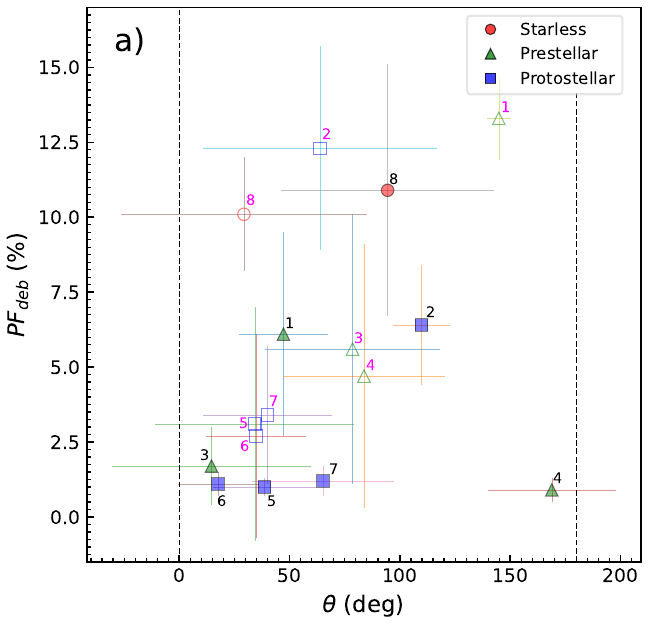}
\label{fig_PFVsBpa}
}
\renewcommand{\thesubfigure}{(b)}
\subfigure
{
%\centering
\includegraphics[width=1\columnwidth]{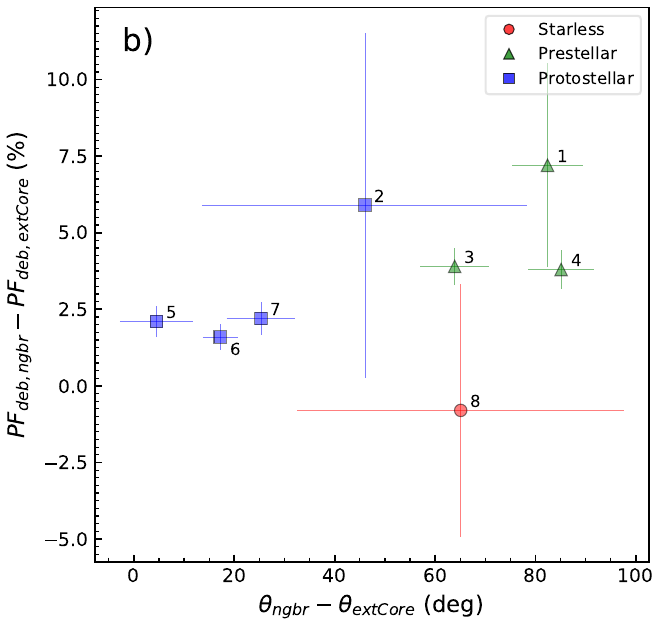}
\label{fig_PFDiffVsBpaDiff}
}
\caption{
\emph{(a)}
\PFdeb\, vs. \bpa\, for the eight extended-core regions (filled symbols) and
the respective neighbourhoods (empty symbols).
The horizontal and vertical extents for each point here show the
($\pm$)standard deviation of the pixels for the region that point represents.
Vertical lines at 0 and 180 degrees
show the extent of the graph, as 180 degree wraps around to 0 degree.
\emph{(b)} Difference of core values from their respective neighbourhood
values. The horizontal and vertical extents for the respective points here
are the error bars.
}
\label{fig_PolParamsComparePlot}
\end{figure}
%\clearpage

Figure \ref{fig_CoreNbrhoodCompare} shows the zoomed-in image of the eight
extended-cores from Figure \ref{fig_quiverPlot_PF}, along with their respective
local neighbourhoods.
The neighbourhood of each extended-core was taken to be a strip of width
2\,pixels (24\arcsec$\sim$\,0.05\,pc)
surrounding a box which fully encompasses all the core pixels.
In the cases where the two cores are close by (such as cores 3 and 4, 
cores 5 and 6), care was taken to exclude the pixels of the nearby core 
from the local neighbourhood.
More details are provided in Table \ref{table_extCoreSummary}.

\subsubsection{Orientation of pixels of cores}
\label{section_PolParams_IndividualCores}

Figure \ref{fig_RosePlotCores} shows the polar histogram of \bpa\, for the
constituent pixels of each core (in yellow) and its neighbourhood (in blue).
The mean values for the cores (yellow line) and neighbourhood (blue line)
are also marked on the plot. It is noteworthy that some of the distributions
are clearly not unimodal. For example, cores 3 and 8 appear bimodal, as do
neighbourhoods of cores 4 and 5. The (circular) mean might not fully capture
the structure in such cases.

We performed the Rayleigh Test of Uniformity for each of the cores using 
the values of their constituent pixels. The null hypothesis posits that the
angles are distributed uniformly around a circle, while the alternate hypothesis
proposes that the angles are not distributed uniformly. Taking the p-value
threshold to be 0.05, we find that we cannot reject the null hypothesis only
for cores 3, 7, and 8. This seems to align with the larger standard
deviation for these cores (see $\delta\theta_{extCore}$ in Table
\ref{table_extCoreSummary} and related discussion in Section
\ref{section_MagneticFieldCalc}); and a visual examination
of half-vectors in Figure \ref{fig_CoreNbrhoodCompare} suggests relatively more
random orientation of pixels for these three cores, especially core 8.
The same test for the neighbourhoods of cores 3-7 (neighbourhoods of
cores 1, 2, and 8 contain too few pixels) finds that the null hypothesis 
cannot be rejected for the neighbourhood of core 5.

To summarize,
while the magnetic field half-vectors of
cores 1, 2, 4, 5, and 6 (and neighbourhoods of cores 3, 4, 6, and 7)
have a definite preferred orientation direction;
the same cannot be said for the magnetic field half-vectors of
cores 3, 7, and 8 (and neighbourhood of core 5), and their pixels could be
randomly oriented.

\subsubsection{Parameters of core regions}
\label{section_PolParams_IndividualCoreParams}

The region contains, in order of evolutionary stages, one starless core (8),
three prestellar cores (1, 3, and 4), and four protostellar cores (2, 5, 6,
and 7).
Figure \ref{fig_PFVsBpa} shows the \PFdeb\, versus \bpa\, plot for each of the
extended-cores (and respective neighbourhoods). The horizontal and vertical
extents for each point here show the standard deviation of the pixels for the
region that point represents.
Note that the actual plotting limits in this figure are 0-180\,degrees, and
the x-axis (i.e. \bpa) at 180\,degree wraps to 0\,degree. The mean polarisation
parameters are also listed in Table \ref{table_extCoreSummary}.

Based on \PFdeb, the region can roughly be classified into an eastern section
with cores 1 and 2, a central dense filamentary section with cores 3-7, and
a western section with core 8.
Cores 1 and 2 on the eastern side appear to have a nearly similar \PFdeb\, of
6.1\% and 6.4\%, respectively, but with clearly different orientations of the
magnetic field. The cores in the central filamentary section show values in a
very narrow range of 0.9-1.7\%. The starless core on the western side has the
highest polarisation fraction of 10.9\%.
In Figure  \ref{fig_PFVsBpa}, \bpa\, shows a similar trend, with the values for
cores 3-6 confined to $\sim$1\,radian range. While core 1 appears to be close
to this cluster of cores 3-6 on the \bpa\, axis, it is clearly differentiated
by its \PFdeb\, value. Similarly, cores 2 and 8 have similar \bpa\, but differ
in \PFdeb. Core 7 also appears to be an outlier on the \bpa\, axis.

\subsubsection{Parameters of neighbourhood regions}
\label{section_PolParams_Neighbourhoods}

In Figure \ref{fig_PolMorphologyCompare}, we see that the cores at the
filament edges -- namely 1, 2, and 8 -- have very few pixels associated with
their neighbourhood.
Apropos of this, it would perhaps be prudent to consider only cores 3-7
as reliable for comparing values with the local neighbourhood. These include
two prestellar (3 and 4) and three protostellar (5, 6 and 7) cores.
\bpa\, for the neighbourhood of these five core regions from east to west are
78\textdegree, 84\textdegree,
34\textdegree, 35\textdegree, and 40\textdegree.
The non-core neighbourhood region thus can be bifurcated into a region
encompassing cores 3 and 4, and another encompassing cores 5, 6, and 7
on the basis of neighbourhood \bpa\, values. This bifurcation also appears
to hold for \PFdeb, as also can be seen in Figure \ref{fig_PFVsBpa}.

On the whole, \PFdeb\, for the neighbourhood appears to decline as one
traverses west from core 3 to 6.
This is in contrast to the values for the cores themselves, which appear
to be stable at $\sim$1$^{+0.7}_{-0.1}$\% for these four cores (3 to 6).
Further west, \PFdeb\, for the neighbourhood increases for core 7.
Even if one
were to consider all the cores, then the trend for the neighbourhood, i.e.
high \PFdeb\, on the eastern side (1 and 2), gradual decline in the central
dense filamentary section (3, 4, 5, and 6), and increase towards the western
side (7 and 8) seems to hold.

\subsubsection{Comparing core parameters with neighbourhood}
\label{section_Core_Neighbourhood_compare}

In this section, we compare the parameters obtained for the individual
extended-cores with the parameters of their respective local neighbourhoods.

Figure \ref{fig_PFDiffVsBpaDiff} shows the difference between the values.
Due to the half-vector nature of the polarisation, the angular difference
is taken as the minima of
($|\bpa_{ngbr}-\bpa_{extCore}|, 180-|\bpa_{ngbr}-\bpa_{extCore}|$),
where \bpa$_{extCore}$ and \bpa$_{ngbr}$ denote the (mean) magnetic field
orientation of the extended-core and its neighbourhood, respectively.
The neighbourhood shows a higher polarisation fraction than the core 
for all except core 8. Though even for core 8, the large error bars and
only a very small deviation preempt any conclusion, and it might well be
that there is no difference in polarisation fraction of this core region
from its neighbourhood.
The stand-alone cores 1 and 2 show large deviation from their neighbourhood
values, which likely is due to poor sampling of their neighbourhood (only 2
pixels each).
Again, considering only core regions 3-7 for our analysis, we find
that prestellar cores appear to have a larger deviation, both in \PFdeb\, and
\bpa, than the protostellar cores.
% Comparison with Planck 2015 XIX Fig 23 ?

Thus overall, there is a clear separation here between the prestellar and
protostellar cores in how they relate to the values of their respective
neighbourhoods.

%%%%%%%%%%%%%%%%%%%%%%%%%%%%%%%%%%%%%%%%

\begin{deluxetable*}{ccccccccccc}[ht]
\digitalasset
\tablewidth{0pt}
\tablecaption{Polarisation Fraction Vs I\label{table_RiceanModel}}
\tablehead{
\colhead{Location}  &
\multicolumn{5}{c}{Ricean\,Mean\,Model} &
\colhead{Null\,Hypothesis} &
\multicolumn{4}{c}{Single-Power\,Law} \\
\cmidrule(lr){2-6}
\cmidrule(lr){7-7}
\cmidrule(lr){8-11}
 &
 Pixels & $\alpha$ & $\sigma_{\textsc{qu}}$ & PF$_{\sigma_{\textsc{qu}}}$
 & $\chi^2_{\mathrm{reduced}}$
 & $\chi^2_{\mathrm{reduced}}$
 & Pixels & $\alpha$ & $\sigma_{\textsc{qu}}$
 & $\chi^2_{\mathrm{reduced}}$
}
\startdata
Filament  & 257 & 0.71$\pm$0.08 & 0.61 & 0.43$\pm$0.16 & 1.08 & 1.42 &
             27 & 1.06$\pm$0.04 & 0.58 & 0.19  \\
Cores     &  96 & 0.85$\pm$0.10 & 0.62 & 1.07$\pm$0.56 & 2.11 & 3.95 &
             27 & 1.01$\pm$0.06 & 0.62 & 0.98  \\
\enddata
\tablecomments{$\sigma_{\textsc{qu}}$ is in units of mJy/beam.
The fitted parameter PF$_{\sigma_{\textsc{qu}}}$ is in fractional terms and
not \%.
}
\end{deluxetable*}

\begin{figure*}
\centering
%
%Section 3.2 from https://cs.brown.edu/about/system/managed/latex/doc/subfigure.pdf
\renewcommand{\thesubfigure}{(a)}
\subfigure
{
%\centering
\includegraphics[width=1\columnwidth]{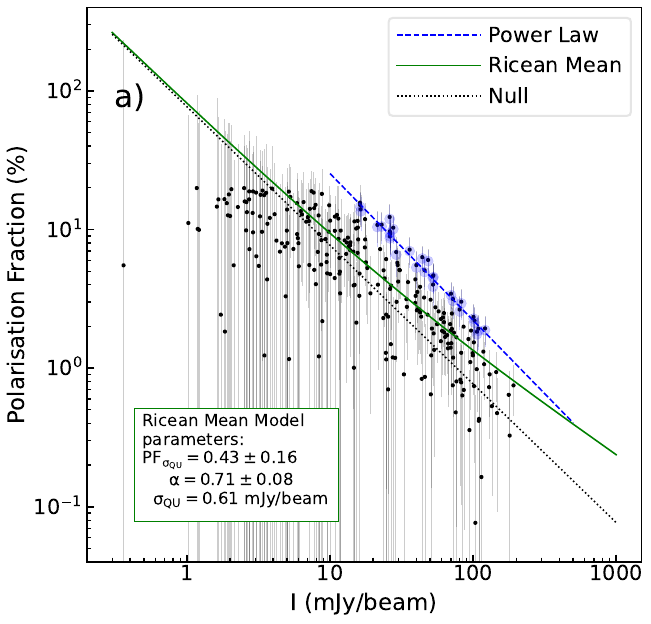}
\label{fig_Ricean_filament}
}
\renewcommand{\thesubfigure}{(b)}
\subfigure
{
%\centering
\includegraphics[width=1\columnwidth]{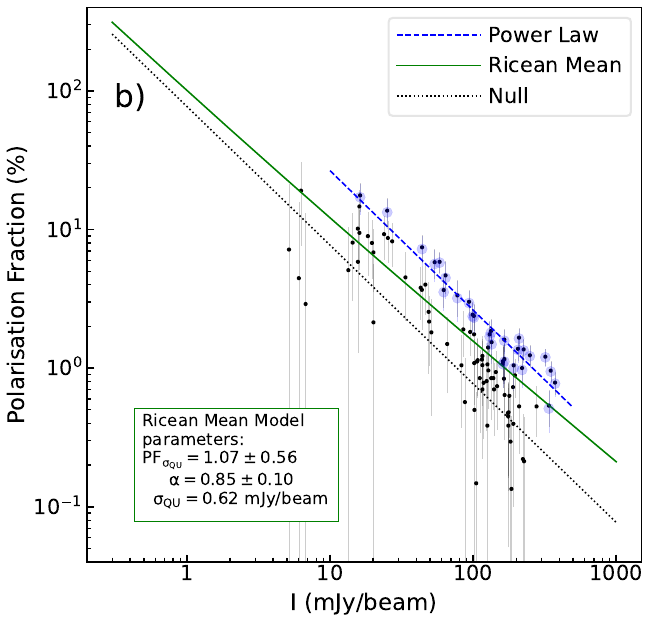}
\label{fig_Ricean_cores}
}
\caption{
\emph{(a)} Polarisation Fraction vs. stokes I relation for all the
filament/non-core pixels.
\emph{(b)} Polarisation Fraction vs. stokes I relation for all the
extended-core pixels.
In both panels,
black dots denote the observed polarisation fractions (\PF) of all the pixels,
i.e. no selection of pixels has been carried out using any (SNR) criteria,
except \PF\,$\leq$20\%. They are used for the Ricean Mean Model fit shown by
the green curve.
Blue circles, on both panels, mark the debiased polarisation fractions (\PFdeb)
of pixels satisfying the high SNR criteria (see Section \ref{section_PFvsI}),
used for single-power law fitting (dashed-blue curve).
Black dashed curve shows the null hypothesis.
The fitted parameters for the Ricean Mean Model are also given.
}
\label{fig_RiceanModel}
\end{figure*}
%\clearpage

\subsection{Polarisation Fraction vs. Total Intensity}
\label{section_PFvsI}

The relationship between the polarisation fraction and total intensity is used
to assess the dust grain alignment efficiency
with respect to
the visual extinction
or optical depth.  As per the theory of radiative torque alignment
\citep{Lazarian_2007MNRAS, Hoang_2008MNRAS, Whittet_2008ApJ, Jones_2015AJ},
dust grain alignment depends on the
wavelength-to-grain size ratio. In regions of high dust extinction, short
wavelength radiation is attenuated leading to a loss in alignment of smaller
grains. The polarisation fraction serves as an indicator of the dust grain
alignment efficiency, with total intensity I acting as a proxy for visual
extinction. This relationship typically shows a power-law dependence of the
form $\PF \propto I^{-\alpha}$, where $0\leq\alpha\leq1$.
The limit $\alpha=0$ gives $\PI \propto I$, which implies that all grains
are equally aligned along each line of sight resulting in a constant
polarisation fraction.
The other limit, $\alpha=1$, implies a constant polarisation intensity
irrespective of the total intensity. Such a case has been interpreted to arise
due to polarised emission from a thin layer of low density material on the
surface of the cloud, overlaying an unpolarised interior
\citep{Pattle_Bistro_Oph_2019ApJ,Kwon_Bistro_SerpensMain_2022ApJ}. The
intermediate values of $\alpha$ imply that polarisation fraction decreases with
increasing intensity and thus signals depolarisation.
This empirical relation is referred to as
\textquotedblleft polarization hole\textquotedblright\, and among its
possible causes are :
loss in grain alignment at high extinction,
complex field geometry within the telescope beam being integrated over
(\textquotedblleft field tangling\textquotedblright)
in the line of sight (LOS), and dust grain
destruction in the vicinity of protostars \citep{Pattle_PPVII_2023ASPC}.

For the calculation of $\alpha$, we use the Ricean Mean Model method of
\citet{Pattle_Bistro_Oph_2019ApJ}
\citep[also see appendix of][]{Pattle_L1689_2021ApJ}. In this method, the
relation between the polarisation fraction and the intensity is
given by :
\begin{equation}
\label{eqn_Rice_PFinitial}
\PF = \PF_{\sigmaQU} \left( \frac{I}{\sigmaQU} \right)^{-\alpha},
\end{equation}
where,
\sigmaQU\, is the rms noise level of the data;
\PF$_{\sigmaQU}$ is the polarisation fraction at this noise level of the
data (i.e. \PF\,(I=\sigmaQU)=\PF$_{\sigmaQU}$); and
$\alpha$ is the power law index in the range $0\leq \alpha \leq 1$.
For a region with N pixels, \sigmaQU\, can be obtained as follows :
\begin{equation}
\sigmaQU = \frac{1}{2N} \displaystyle\sum_{i=1}^{N} \left( \sqrt{\sigmaQi^2} + \sqrt{\sigmaUi^2} \right),
\end{equation}
where $\sigmaQi^2$ and $\sigmaUi^2$ represent the variances of the \emph{i-th}
pixel. \sigmaQU\, is thus a property of the dataset, and the aim is to obtain
\PF$_{\sigmaQU}$ and $\alpha$.

Since polarisation fractions are Rice distributed, the mean of the Rice
distribution gives us \citep[see][for details]{Pattle_Bistro_Oph_2019ApJ} :
\begin{equation}
\label{eqn_Rice_PF_RiceanMean}
%\footnotesize
\PF = \sqrt{\frac{\pi}{2}}\!\left(\!\frac{I}{\sigmaQU}\!\right)^{\!-\!1}\!\mathcal{L}_{_{\frac{1}{2}}}\!\!\left(\frac{-\PF_{\sigmaQU}^{\,2}}{2}\!\left(\!\frac{I}{\sigmaQU}\!\right)^{\!2(1-\alpha)} \right),
\end{equation}
where $\mathcal{L}_{1/2}$ is the Laguerre polynomial of order 1/2. Apart from
equation \ref{eqn_Rice_PFinitial}, the above equation also assumes
$\sigma_{\!_{\PF}}$=\sigmaQU/I.
We fit Equation \ref{eqn_Rice_PF_RiceanMean} using the observed (i.e.
non-debiased) polarisation fraction and the total intensity I, keeping
\PF$_{\sigmaQU}$ and $\alpha$ as free parameters. This fit was carried out
using all the pixels in our region of analysis, i.e. no trimming of dataset
was carried out on the basis of SNR criteria on I or \PF.
Only \PF\,$\leq$20\%
%and RA-Dec selection criteria were
was applied, as pixels with \PF\,$>$20\% are likely unphysical (see
Section \ref{section_PolParams}).
The fitting was carried out using the \texttt{scipy} routine \texttt{curve\_fit}.
The error in \PF\, was also taken into account for the fit.
Figure \ref{fig_RiceanModel} shows the model fit for the filament and the
extended-core regions separately. Solid green line denotes the Ricean Mean
Model fit. The obtained parameters are given in Table \ref{table_RiceanModel}.

To compare with analyses in literature,
the figures also show
the null hypothesis (dotted line), and
the single-power law (blue dashed line).
For the null hypothesis, \PF$_{\sigmaQU}$=0, and thus
\PF\,$=\sqrt{\pi/2} \left(I/\sigmaQU \right)^{-1}$ for all I.

The single-power law is fit is carried out using \PFdeb\, and using only the
pixels which satisfy a high SNR criteria. A high SNR criteria is used as
the polarisation fraction tends to a Gaussian distribution in the high SNR
limit, and in a low SNR limit, the power-law index of $-$1 will always be
obtained.
Debiased values are used as not doing so would artificially inflate $\alpha$.
We refer to \citet{Pattle_Bistro_Oph_2019ApJ} for a detailed discussion of
the above.
The fit is calculated using the equation
\citep{Pattle_Bistro_Oph_2019ApJ} :
\begin{equation}
\PFdeb = PF_{\sigmaQU} \left( \frac{I}{\sigmaQU} \right)^{-\alpha}.
\end{equation}
Only the pixels which satisfied the (commonly taken) high SNR criteria --
SNR(\PFdeb)$>$3, SNR(I)$>$10, and $\sigma_{\scriptscriptstyle \PFdeb}\!<5\%$
-- were utilised.

The reduced $\chi^2$ for is calculated as follows (given \emph{n} datapoints
used for fitting) for the two models :
\begin{eqnarray}
\chi^2_{\mathrm{reduced}} = \frac{ \chi^2}{n-2} = \frac{\displaystyle\sum_{i=1}^{n} \left( \frac{\PF_{model}-\PF_i}{\sigma_{_{\PF_i}}} \right)^2}{n-2}.
\end{eqnarray}
The same has also been reported in Table \ref{table_RiceanModel} to assess
the goodness-of-fit for each curve.
The reduced $\chi^2$ for the null hypothesis is given by
$\chi^2_{\mathrm{reduced}}=\chi^2/n$, and it was calculated using
the same pixels and values as for the Ricean Mean Model (i.e. observed).

%%%%%%%%%%%%%%%%%%%%%%%%%%%%%%%%%%%%%%%%

\subsection{Histogram of Relative Orientations}
\label{section_HRO}

The HRO (Histogram of Relative Orientations) is used to assess the relative
orientation of the magnetic field vectors with respect to the column density
structures \citep{Soler_HRO_2013ApJ}. We adopt the steps for calculation from
\citet{Planck_XXXV_2016AA}. A brief description of the procedure follows.
The angle between the tangent to the iso-column density contours and the
\Bpos\, vector can be given by :
\begin{eqnarray}
\label{eqn_HRO_formula}
\phi &=& arctan \frac{|\bm{\nabla \NHtwo} \times \bm{E}|}{\bm{\nabla \NHtwo} . \bm{E}}
\\
     &=& arctan \frac{|\bm{\nabla \NHtwo}| |\bm{E}| sin\phi}{|\bm{\nabla \NHtwo}| |\bm{E}| cos\phi},
\end{eqnarray}
where,
$\bm{\nabla \NHtwo}$ is the gradient of the column density map, and $\bm{E}$
is the polarisation vector.
The gradient is perpendicular to the iso-\NHtwo\, contours and the polarisation
vector ($\bm{E}$) is perpendicular to the \Bpos\, vector in FIR. Hence $\phi$
-- the angular difference between the gradient and the polarisation --
gives us the angular difference of the tangent to the column density contour
and magnetic field vector.
The gradient is given by the following convolution operation
(denoted by $\otimes$) :
\begin{eqnarray}
\label{eqn_gradient}
\bm{\nabla \NHtwo} &=& (G_x^l \otimes \NHtwo) \bm{\hat{x}} + (G_y^l \otimes \NHtwo) \bm{\hat{y}} \\
              &=& g_x \bm{\hat{x}} + g_y \bm{\hat{y}},
\end{eqnarray}
where, G$_x^l$ and G$_y^l$ are the gaussian derivative kernels (of kernel size
$l \times l$) along the x and y-axis, respectively. The orientation angle of
the gradient is then given by :
\begin{equation}
\label{eqn_psi}
\psi = arctan(-g_x/g_y).
\end{equation}
Application of the convolution operation given in Equation \ref{eqn_gradient} on
the column density map gives us the (pixel-by-pixel) x- and y-components of the
gradient map, which are then used to obtain a map of the gradient angle
$\psi$ using Equation \ref{eqn_psi}. We utilized the \texttt{scipy} task
\texttt{gaussian\_filter} with a kernel size of 3\,pixel$\times$3\,pixel for
our purpose.
The pixel-by-pixel (relative) difference of the gradient angle map ($\psi$)
and the polarisation angle map ($\theta$) gives us relative orientation
($\phi$) map.
Due to the half-vector nature of the polarisation vector, the difference
between the gradient and the polarisation
angle has to be mapped to a 0-90\textdegree\, interval. Also, since we are only
concerned with the relative orientation between the polarisation (half-)vector
and the gradient vector, $\phi$ is physically equivalent to $-\phi$, and the
angular difference can be mapped to (0,90) rather than (-90,90) without any
loss of generality \citep{Soler_HRO_2017AA}. For example, $\phi=0$ indicates
that the magnetic field is parallel to the (tangent of the) column density
contour, and $\phi=\pm90$ indicates that it is perpendicular.
The pixel-by-pixel map of $\phi$ is then used to construct the histogram.

\begin{figure}
\centering
\includegraphics[width=\columnwidth]{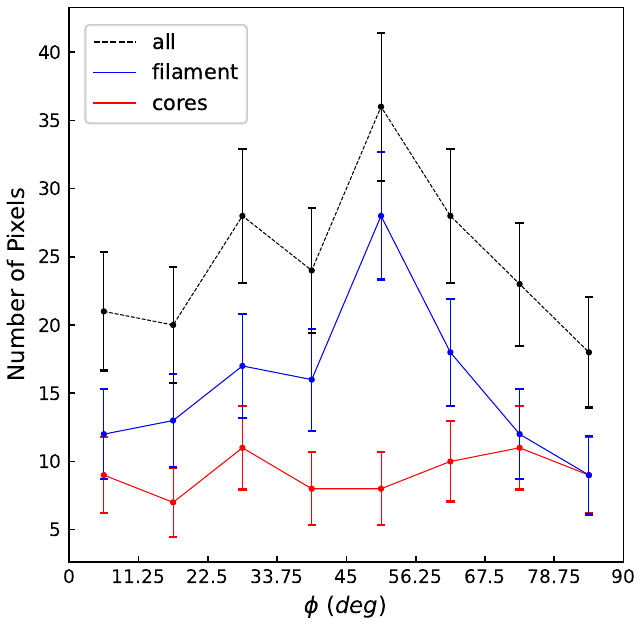}
\caption{
Histogram of Relative Orientations for extended-cores (red),
filament (blue), and all the pixels (black-dashed).
}
\label{fig_HRO_byType}
\end{figure}

Figure \ref{fig_HRO_byType} shows the HROs for the filament (in blue) and the
extended-cores (in red). The HRO for all the pixels is also shown by dashed
black line.
Error $\sigma_k$ for the \emph{k-th} bin was calculated using the following
\citep{Planck_XXXV_2016AA} :
\begin{eqnarray}
\sigma_k^2 = h_k \left( 1 - \frac{h_k}{h_{tot}} \right),
\end{eqnarray}
where $h_k$ is the count of the \emph{k-th} bin, and $h_{tot}$ is the total
count.

The HRO shows two peaks for the overall region - first peak in the
22.5 to 33.75\,degree bin and a second peak in the 45 to 56.25\,degree bin.
The second peak is clearly due to the orientation of the filament pixels,
as is also seen in the blue histogram. The first peak appears to be due to
the combined effect of core and filament pixels, and thus it can be surmised
that core pixels are partly aligned with the filament pixels.

%TBD :
%Wang NGC 2264 - projected rayleigh statistic;
%Rayleigh statistic as in Jow 2018 paper.

%%%%%%%%%%%%%%%%%%%%%%%%%%%%%%%%%%%%%%%%

\begin{figure*}[ht]
\centering
%
%Section 3.2 from https://cs.brown.edu/about/system/managed/latex/doc/subfigure.pdf
\renewcommand{\thesubfigure}{(a)}
\subfigure
{
%\centering
\includegraphics[width=\columnwidth]{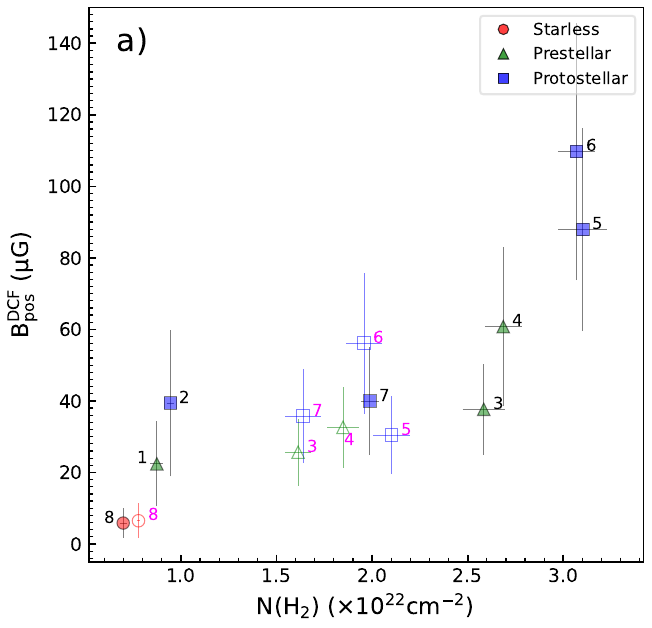}
\label{fig_BposVsNH2_DCF}
}
\renewcommand{\thesubfigure}{(b)}
\subfigure
{
%\centering
\includegraphics[width=\columnwidth]{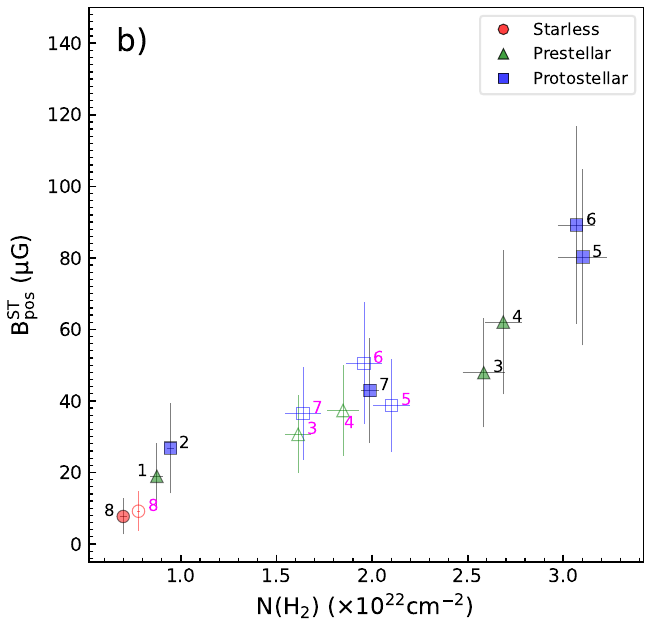}
\label{fig_BposVsNH2_ST}
}
\caption{
Plane-of-sky magnetic field strength (\Bpos) as a function of the mean column
density for the extended-cores (filled symbols) and the respective neighbourhoods
(empty symbols).
\emph{(a)} using DCF method.
\emph{(b)} assuming compressible ISM \citep{SkalidisTassis_2021AA}.
}
\label{fig_BposVsNH2}
\end{figure*}
%\clearpage

\subsection{Magnetic Field using DCF}
\label{section_MagneticFieldCalc}

\subsubsection{Calculation}
\label{section_MagneticFieldCalc_DCFCalculation}

We utilize the widely used Davis-Chandrasekhar-Fermi (DCF) formalism
\citep{Davis_1951PhRv, ChFermi_1953ApJ} to calculate the magnetic field
strength in the extended-core regions and the associated neighbourhoods. 
The DCF method works on the assumption that over and
above the underlying large-scale mean field direction, there are perturbations
in the magnetic field on small-scale due to the non-thermal motions of the gas.
The turbulence generated by this non-thermal motion is Alfvenic in nature.
The dispersion in the deviation of polarisation angles from the mean field
direction can be used to calculate the plane-of-sky magnetic field (\Bpos)
using the formula \citep{Crutcher_2004ApJ} :
\begin{equation}
\label{eqn_Bpos}
\BposDCF \approx 9.3\,\frac{\sqrt{\nHtwo}~~\Delta\!\mathrm{V_{NT}}}{\delta \theta}
\end{equation}
where
\nHtwo\, is the number density of hydrogen molecules (in cm$^{-3}$),
$\Delta\!\mathrm{V_{NT}}$ is the linewidth (\textsc{fwhm}) of non-thermal
velocity component (in \kmps), and
$\delta \theta$ is the dispersion in polarisation angles (in degrees).
The resultant \BposDCF\, is in microgauss.

The non-thermal velocity linewidth is calculated using the expression
\citep{FullerMyers_1992ApJ, FiegePudritz_2000MNRAS} :
\begin{equation}
\Delta\!\mathrm{V_{NT}} = \sqrt{(8\,ln2) \times (\sigma_{tot}^2-c_s^2)}
\end{equation}
where $\sigma_{tot}$ is the total velocity dispersion,
c$_s$($=\sqrt{kT/\bar{m}}$) is the speed of sound,
k is the Boltzmann constant,
T is the temperature, and
$\bar{m}$ is the mean molecular weight of the medium.

Taking T as 10\,K and $\bar{m}$ as 2.33\,amu \citep{Kauffmann_2008AA}, we
calculated the sound speed to be $\sim$0.19\,\kmps.
$\sigma_{tot}$ was calculated using the \NHtwo\, vs. $\sigma_{tot}$ plot from
\citet[][using N$_2$H$^+$\,(J=1-0), which is a good tracer of dense
gas; see their Figure 6]{Arzoumanian_2013AA}.
$\Delta\!\mathrm{V_{NT}}$ was found to lie in the range 0.2-0.7\,\kmps,
and seems to be consistent with other core and filament studies in
literature
\citep{Kwon_Bistro_OphA_2018ApJ,Kwon_Bistro_SerpensMain_2022ApJ,
Soam_Bistro_OphB_2018ApJ,Wu_Bistro_ISF_2024ApJ}.
The number density (\nHtwo) was calculated by dividing the column density
value by $W_{fil}=$0.1\,pc, which has been measured to be the width of
the filament
\citep{Arzoumanian_OriBupfil_2017MmSAI, Arzoumanian_IC5146_2011AA,
Andre_NGC6334_2016AA}.
Thus, for each of the extended-core regions, we first obtained the mean 
\NHtwo\, using their constituent pixels in the column density map, which 
was then used to calculate \nHtwo\, and $\Delta\!\mathrm{V_{NT}}$.
Archival N$_2$H$^+$\,(J=1-0) observations of \object{PGCC G203.21-11.20}
\citep{Yi_LambdaOrionisIII_2021ApJS}, which
encompass core 3 to 6 regions, show the non-thermal linewidth to be in the
range $\sim$0.6-0.7\,\kmps. Thus our estimates (for these four cores) are not
only consistent with observations,  but also the error in
$\Delta\!\mathrm{V_{NT}}$ considered for further calculation
($\sim$0.12\,\kmps, see Appendix \ref{appendix_Errors_Bpos})
takes into account the range of variation.
For angular dispersion, the circular standard deviation of the pixel angles
was used as a measure.

Subsequently, for each of the extended-cores in Figure \ref{fig_quiverPlot_PF},
\BposDCF\, was calculated using Equation \ref{eqn_Bpos}. The calculated values
are tabulated in Table \ref{table_extCoreSummary} and graphically
shown in Figure \ref{fig_BposVsNH2}. Error on \Bpos\, was calculated using
error propagation of the errors on column density
($\sim$0.1-1$\times$10$^{21}$\,cm$^{-2}$), filament width ($\sim$0.05\,pc),
non-thermal velocity \textsc{fwhm} ($\sim$0.12\,\kmps), and
angular dispersion ($\sim$3-30\,degrees),
and is discussed in Appendix\,\ref{appendix_Errors_Bpos}.
Same procedure was used for the calculation of the \Bpos\, values of the
neighbourhood of the cores, except for those of cores 1 and 2, as these
neighbourhoods contain only 2 pixels each (see Table \ref{table_extCoreSummary}
and Figure \ref{fig_PolMorphologyCompare}), and thus a reliable measure of
angular dispersion cannot be obtained for them.

\subsubsection{Analysis of DCF Field Strengths}
\label{section_DCFresults}

Figure \ref{fig_BposVsNH2_DCF} shows the variation of magnetic field strength
with column density for the different extended-cores (as filled symbols).
The magnetic field values for the neighbourhood of cores has also been plotted 
(as unfilled symbols) and labelled.

Prima facie for the cores, there is a clear trend of increase in \Bpos\, 
with increasing \NHtwo. However, different regimes in \NHtwo\, show differing
trends.
In the low-density regime -- $\sim$0.6-1$\times$10$^{22}$cm$^{-2}$ -- lie the
cores at the edges of the filament (cores 1, 2 and 8).
Here, the protostellar core 2 shows the highest magnetic field strength. 
The starless core 8 has the lowest field strength among all the cores in our
analysis, and it is close to the value of its surrounding neighbourhood.
In the intermediate range $\sim$1-2$\times$10$^{22}$cm$^{-2}$, the protostellar
cores 2 and 7 show remarkably similar \Bpos\, despite \NHtwo\, differing by a
factor of $\sim$2. The \Bpos\, values for the neighbourhood also seem to be
clustered in this regime.
In the high-density part of the filament, corresponding to
\NHtwo$\sim$2.5-3$\times$10$^{22}$cm$^{-2}$, there is a steep rise in the
magnetic field strength. The protostellar core 6 shows the highest \Bpos\,.
Two out of three prestellar cores lie in this regime.
Comparison with literature shows our values to be consistent with that
calculated for regions with similar column density in other molecular
clouds \citep{Eswaraiah_Bistro_Taurus_2021ApJ,Konyves_Bistro_Rosette_2021ApJ,
Kwon_Bistro_SerpensMain_2022ApJ}.

If examined on the basis of cores, then for protostellar cores, \Bpos\,
-- which remained roughly stable as \NHtwo\, increased from $\sim$1 to
2$\times$10$^{22}$cm$^{-2}$ -- shows a sharp increase when going from $\sim$2
to 3$\times$10$^{22}$cm$^{-2}$. Prestellar cores also show low field strength
in the low-density regime, with field strength increasing sharply in the
high-density regime. We lack prestellar data points in the intermediate
density regime to fully determine whether the value remains stable -- like it 
does for protostellar cores -- or not. Nevertheless, the trend from core 1 to 
core 3 does suggest a much gentler slope than that from core 3 to core 4.
At the low-density regime and the high-density regimes, where we can compare
magnetic field strengths of different types of cores, we find that the
protostellar core appears to show the highest magnetic field strength when
compared with other core types.

\subsubsection{Sources of Uncertainty in the DCF method}
\label{section_DCFIssues}

The classical DCF method is expected to give reliable estimations of the
magnetic field for $\delta \theta <$25\textdegree\,
\citep{Ostriker_2001ApJ,Crutcher_2012ARAA}. Though this is not so for
cores 3, 4, 7, and 8; the departure from 25\textdegree\, is significant
for 3 and 8, making their magnetic field strength estimations less robust
than that for the rest of the core regions.
Similarly, for the neighbourhoods, except for core 6, all others were found
to possess large angular dispersions.

The DCF method assumes an equipartition
of the turbulent kinetic energy and the turbulent magnetic energy.
While the assumption is expected to be satisfied in the sub-Alfvenic regions;
for
super-Alfvenic regions, the magnetic field perturbations cannot keep up with
the strong turbulent motions and the assumption breaks down
\citep{FalcetaGoncalves_2008ApJ}. Using DCF in such a case would lead to
an overestimation of the magnetic field strength
\citep{Liu_NumSim_2021ApJ,Liu_2022FrASS}. According to our calculation later
in the paper (in Section \ref{section_EnergyBalance} and
Table \ref{table_CoreSummary}), cores 3 and 4 are likely to be super-Alfvenic,
and hence their field strengths could be overestimated.

Finally, we note that the efficacy of DCF for estimating field strengths of
protostellar objects could be affected by phenomena such as outflows and
gravitational collapse. However, any detailed modelling to quantify these
effects is outside the scope of this paper.

%%%%%%%%%%%%%%%%%%%%%%%%%%%%%%%%%%%%%%%%

\subsection{Magnetic Field assuming compressible ISM}
\label{section_MagneticFieldCalc_ST}

The classical DCF method has the assumption of incompressible fluids
in its calculation. However, the ISM is compressible and hence the field
strength should be calculated using compressible turbulence.
While DCF (Section \ref{section_MagneticFieldCalc}) is the mainstay for
further calculations in the paper, we estimate the field strength assuming
compressible turbulence for comparison.
We use the method provided by \citet{SkalidisTassis_2021AA} for this purpose.
The following expression from \citet{Choi_IC348_2024ApJ} was used to calculate
the magnetic field :
\begin{equation}
\label{eqn_Bpos_ST}
\BposST \approx 1.76\,\frac{\sqrt{\nHtwo}~~\Delta\!\mathrm{V_{NT}}}{\sqrt{\delta \theta}}
\end{equation}
where the symbols on the right hand side retain the same meaning as in Equation
\ref{eqn_Bpos}, and \BposST\, is the field strength in microgauss.
The values for the extended-cores and their neighbourhoods are
also listed in Table\,\ref{table_extCoreSummary}. While the change
in \Bpos\, is well within the error limits, no clear pattern can be discerned
whether the \citet{SkalidisTassis_2021AA} method leads to a larger or smaller
field strength overall.

Figure\,\ref{fig_BposVsNH2_ST} shows the field strength
variation with column density. Despite a quantitative change in strengths, 
the figure shows that that the relative trend between the cores remains almost 
same. 
In both the low-density ($\sim$0.6-1$\times$10$^{22}$cm$^{-2}$) and
high-density regimes ($\sim$2.5-3$\times$10$^{22}$cm$^{-2}$),
the increase in \Bpos\, is less steep than that for DCF method.
In the intermediate range ($\sim$1-2$\times$10$^{22}$cm$^{-2}$), the field
strength does not appear to be nearly constant for this case though, and there
is a increase in \Bpos\, with \NHtwo.
This method also contains uncertainties, and we refer to
\citet{Liu_2022FrASS,Lazarian_2022ApJ,Li_B211_2022MNRAS} for a detailed
discussion on the method and its shortcomings.

%%%%%%%%%%%%%%%%%%%%%%%%%%%%%%%%%%%%%%%%%%%%%%%%%%%%%%%%%%%%%%%%%%%%%%%%%%%%%%%%

\section{Implications of polarisation results for Filament and Core dynamics}
\label{section_results_FilamentAndCores}

%%%%%%%%%%%%%%%%%%%%%%%%%%%%%%%%%%%%%%%%

\subsection{Criticality of the Filament}
\label{section_Criticality}

We calculate the maximum line-mass for a magnetized filament using the
following expression
\citep{Tomisaka_2014ApJ,Hanawa_2015ApJ,Kashiwagi_2021ApJ}:
\begin{equation}
\label{eqn_MBlinecrit}
\small
M^B_{_{\substack{line,\\ crit}}} \!\! \simeq \! \left[ \! 2.24\! \left( \frac{R_o}{0.05\,\mathrm{pc}} \right) \left( \! \frac{B_o}{10\,\mu G}\! \right) \! \right]\msun \mathrm{pc}^{-1} + 1.66 \frac{\sigma_{{tot}}^2}{G}
\end{equation}
where
M$^B_{line,crit}$ is the maximum line-mass (calculated in \msun\,pc$^{-1}$);
R$_\mathrm{o}$ is the radius of the filament;
B$_\mathrm{o}$ is the magnetic field in the filament; and
$\sigma_{{tot}}$ is the total velocity dispersion.
Since the width of the filament is taken as 0.1\,pc, hence we take
R$_\mathrm{o}$ as 0.05\,pc. We take B$_\mathrm{o}$ to be the average of
\Bpos\, values for the neighbourhoods of the cores
(B$^{\textsc{dcf}}_{pos,ngbr}$ in
Table \ref{table_extCoreSummary}), which comes out to be
$\sim$31.2\,$\text{\textmu}$G.
Using the mean \NHtwo\, of the contiguous filamentary structure detected
above 5\,$\sigma$ in stokes\,I image
(see Appendix \ref{appendix_radialVariation} and Figure \ref{fig_CrestLine}),
$\sigma_{{tot}}$ was determined to be $\sim$0.28\,\kmps\, from the
\NHtwo\, vs. $\sigma_{tot}$ relation of \citet{Arzoumanian_2013AA}.
Thus, M$^B_{line,crit}$ for the filament is calculated to be
$\sim$37\,\msun\,pc$^{-1}$.
Taking the transverse line profile of the filament along the north-south
direction in the column density map, we obtained the transverse extent of
the filament, and subsequently the total mass of the filament as
$\sim$\,100\msun. The resulting line mass, taking the length to be
$\sim$\,1.25\,pc, comes out to be 80\,\msun\,pc$^{-1}$. Thus the filament
is thermally and magnetically supercritical.

%%%%%%%%%%%%%%%%%%%%%%%%%%%%%%%%%%%%%%%%

\begin{deluxetable*}{cp{2.75cm}|r|rrr|rrrr}[ht]
%\tabletypesize{\scriptsize}
\digitalasset
\tablewidth{0pt}
\tablecaption{Estimates of values for the cores \label{table_CoreSummary}}
\tablehead{
\colhead{S No.}  & \colhead{Quantity}  &
\colhead{Core 8} & \colhead{Core 1} & \colhead{Core 3} & \colhead{Core 4} &
\colhead{Core 2} & \colhead{Core 5} & \colhead{Core 6} & \colhead{Core 7}
}
\startdata
\hline
  1 &                                           CoreType &        Starless &      Prestellar &      Prestellar &      Prestellar &    Protostellar &    Protostellar &    Protostellar &    Protostellar  \\
  2 &                                    Mass\,$(\msun)$ &   0.13$\pm$0.06 &   0.97$\pm$0.33 &   8.91$\pm$1.80 &   9.13$\pm$2.02 &   0.29$\pm$0.10 &   1.84$\pm$0.92 &   4.82$\pm$2.41 &   1.22$\pm$0.61  \\
  3 &                        \nHtwo\,(10$^4$\,cm$^{-3}$) &             0.3 &             3.0 &             7.0 &             9.6 &             1.0 &             7.2 &            19.4 &             3.0  \\
\hline
%\cmidrule(lr){1-2}
\multicolumn{7}{l}{Energy Parameters of Cores} \\
\cmidrule(lr){1-2}
  4 &                     $\lambda^{\textsc{dcf}}_{obs}$ &     0.8$\pm$0.5 &     2.0$\pm$1.1 &     4.5$\pm$1.5 &     3.5$\pm$1.3 &     0.4$\pm$0.2 &     1.2$\pm$0.4 &     2.5$\pm$0.8 &     1.2$\pm$0.5  \\
  5 &                      $\lambda^{\textsc{st}}_{obs}$ &     0.6$\pm$0.4 &     2.4$\pm$1.2 &     3.5$\pm$1.1 &     3.4$\pm$1.1 &     0.5$\pm$0.3 &     1.3$\pm$0.4 &     3.0$\pm$0.9 &     1.1$\pm$0.4  \\
  6 &              V$_{\mathrm{A,core}}$\,(km\,s$^{-1}$) &            0.14 &            0.17 &            0.19 &            0.26 &            0.52 &            0.43 &            0.33 &            0.30  \\
  7 &                              M$_{\mathrm{A,core}}$ &            0.61 &            0.72 &            1.44 &            1.07 &            0.26 &            0.69 &            0.90 &            0.77  \\
  8 &          E$_{\mathrm{KE,NT,core}}$\,(10$^{35}$\,J) &            0.03 &            0.43 &           19.13 &           20.41 &            0.15 &            4.76 &           12.33 &            1.96  \\
  9 &           E$_{\mathrm{KE,T,core}}$\,(10$^{35}$\,J) &            0.14 &            1.03 &            9.48 &            9.72 &            0.31 &            1.96 &            5.13 &            1.30  \\
 10 &              E$_{\mathrm{B,core}}$\,(10$^{35}$\,J) &            0.03 &            0.28 &            3.08 &            5.98 &            0.77 &            3.36 &            5.10 &            1.10  \\
 11 &                   E$_{\mathrm{G}}$\,(10$^{35}$\,J) &           -0.02 &           -1.10 &          -59.04 &          -67.89 &           -0.10 &           -4.34 &          -29.80 &           -1.62  \\
 12 &           E$_{\mathrm{B}}$/E$_{\mathrm{KE,total}}$ &            0.15 &            0.19 &            0.11 &            0.20 &            1.66 &            0.50 &            0.29 &            0.34  \\
\hline
\multicolumn{7}{l}{Energy Parameters of Neighbourhoods} \\
\cmidrule(lr){1-3}
 13 &              V$_{\mathrm{A,ngbr}}$\,(km\,s$^{-1}$) &            0.05 &               - &            0.15 &            0.17 &               - &            0.15 &            0.29 &            0.20  \\
 14 &                              M$_{\mathrm{A,ngbr}}$ &            1.94 &               - &            1.39 &            1.28 &               - &            1.57 &            0.79 &            1.02  \\
 15 &          E$_{\mathrm{KE,NT,ngbr}}$\,(10$^{35}$\,J) &            0.12 &               - &            7.99 &            9.87 &               - &           10.87 &           11.19 &            5.27  \\
 16 &           E$_{\mathrm{KE,T,ngbr}}$\,(10$^{35}$\,J) &            0.40 &               - &            6.83 &            7.12 &               - &            6.74 &            7.54 &            4.42  \\
 17 &              E$_{\mathrm{B,ngbr}}$\,(10$^{35}$\,J) &            0.01 &               - &            1.38 &            2.02 &               - &            1.47 &            5.99 &            1.70  \\
 18 &           E$_{\mathrm{B}}$/E$_{\mathrm{KE,total}}$ &            0.02 &               - &            0.09 &            0.12 &               - &            0.08 &            0.32 &            0.18  \\
\enddata
\tablecomments{\emph{(Row-wise)}
1 - Type of core (see Table\,\ref{table_extCoreSummary}).
2 - Mass of the core from the HGBS catalog.
3 - Number density of the core from the HGBS catalog.
4,5 - Mass-to-flux ratio calculated using \BposDCF\, and \BposST\, from
Table\,\ref{table_extCoreSummary}, respectively.
6 - Alfven velocity of the core.
7 - Alfvenic Mach number of the core.
8 - Non-thermal kinetic energy of the core.
9 - Thermal kinetic energy of the core.
10 - Magnetic energy of the core.
11 - Gravitational potential energy of the core.
12 - Ratio of magnetic energy to total KE of the core.
13 to 17 - Same as rows 6 to 10, but for the local neighbourhood.
18 - Same as row 12, but for the local neighbourhood.
For cores 1 and 2, the values for the neighbourhood cannot be calculated as
an estimate of magnetic field strength could not be obtained for them
(see Table \ref{table_extCoreSummary}).
}
\end{deluxetable*}

\subsection{Stability of the Cores}
\label{section_CriticalityCores}

In this section, we analyse the mass-to-flux ratio, the different energy terms
and associated physical quantities of the individual
HGBS \citep{Konyves_2020AA_HGBS} cores to assess their stability.
For this calculation, core parameters like mass, effective radius, column
density, and volume number density tabulated in the HGBS catalog are used.
We assume that the plane of sky magnetic field strength (\Bpos)
for each HGBS core to be same as that calculated for the extended-core region
in Section\,\ref{section_MagneticFieldCalc}.

%%%%%%%%%%%%%%%%%%%%%%%%%%%%%%%%%%%%%%%%

\subsubsection{Mass-to-Flux Ratio}
\label{section_MassToFluxRatio}

The mass (M) to flux ($\Phi$) ratio ($\lambda$) in terms of critical value of
(1/2$\pi\sqrt{G}$) is given by
\citep{NakanoNakamura_MassToFluxRatio_1978PASJ, Crutcher_2004ApJ} :
\begin{eqnarray}
\label{eqn_lambda}
\lambda_{obs} &=& 7.6\times10^{-21} \left( \frac{\NHtwo}{B} \right),
\end{eqnarray}
where,
\NHtwo\, is the column density of the core (in cm$^{-2}$) from the HGBS
catalog, and
B is the magnetic field (in $\text{\textmu}$G)
calculated in Section\,\ref{section_MagneticFieldCalc}.

We calculate $\lambda_{obs}$ assuming B=\Bpos\, and using both
\BposDCF and \BposST.
The results are given in Table \ref{table_CoreSummary}.
While the protostellar cores (2, 5, 6, and 7) span the domain from
subcritical (core 2, $\lambda_{obs} < 1$) to
transcritical (core 5 and 7, $1 \lesssim \lambda_{obs} \lesssim 2$) to
supercritical (core 6, $\lambda_{obs} \gtrsim 2$);
all the prestellar cores (1, 3, and 4) are found to be supercritical; and
the sole starless core (core 8) is subcritical.
A transcritical or supercritical core implies that magnetic field support
is not sufficient to resist against gravity.

If one were to utilise the total field strength $|\textbf{B}|$ instead of
\Bpos\, -- given by
$|$\textbf{B}$|$=($4/\pi$)\Bpos \citep{Crutcher_2004ApJ} --
in Equation \ref{eqn_lambda},
as well as apply the statistical correction due to the orientation of the
cores, i.e. $\lambda=\lambda_{obs}/3$ \citep{Crutcher_2004ApJ},
then all cores except 3 and 4 become subcritical.
The prestellar cores 3 and 4 can be said to be transcritical in such a case,
with $\lambda^{\textsc{dcf}} \sim 1.2$ and $\sim 0.9$, respectively.

\subsubsection{Energy Balance}
\label{section_EnergyBalance}

In this section, we look at the different energy terms. It should be noted that
the energy values calculated here contain significant uncertainties, and should
be treated as a representative of their order of magnitude.
The Alfven velocity of each of the cores was calculated using the following
expression (in cgs system) :
\begin{equation}
\label{eqn_Velocity_Alfven}
V_A = B/\sqrt{4\pi \rho} = B/\sqrt{4\pi (\muHtwo \mathrm{m_{H}} \nHtwo)},
\end{equation}
% Refs : Hwang2021-OMC1, Karoly2023-L43, Eswaraiah2021-Taurus
where
\muHtwo(=2.8) is the molecular weight per hydrogen molecule
\citep{Kauffmann_2008AA},
$\mathrm{m_H}$ is the mass of hydrogen atom,
B is the magnetic field strength in Gauss,
$\rho$ is the density (in g\,cm$^{-3}$), and
\nHtwo\, is the volume number density of the region.
Subsequently, the Alfvenic Mach number is given by :
\begin{equation}
M_A = \sigma_{NT}/V_A,
\end{equation}
%ref : Wang2019-IC5146
where $\sigma_{NT}$ for each core is that calculated
%from \citet{Arzoumanian_2013AA}
in Section \ref{section_MagneticFieldCalc}.
The magnetic energy term can be then obtained as:
\begin{eqnarray}
E_B &=& \frac{B^2}{8 \pi}\,V = \frac{1}{2}m V_A^2,
\end{eqnarray}
where,
B and V are the magnetic field strength and volume (of the core) in cgs
units, respectively; and
$m$ is the mass of the core.
We use the right hand side expression with V$_A$ for calculation.
The total kinetic energy (KE) term (E$_{KE}$) can be expressed as a sum of
thermal KE (E$_{_{KE,T}}$) and non-thermal KE (E$_{_{KE,NT}}$) as :
\begin{eqnarray}
E_{_{KE}}\!=\!\frac{3}{2}m\,\sigma_{_{total}}^2 &=& \frac{3}{2}m\,c_s^2\!+\!\frac{3}{2}m\,\sigma_{_{NT}}^2 \\
                                               &=& E_{_{KE,T}}\!+\!E_{_{KE,NT}}.
\end{eqnarray}
Assuming the cores to be spheres with effective radius R
(\textquoteleft observed core radius\textquoteright\, from the HGBS catalog),
the gravitational (potential) energy (E$_G$) is :
\begin{eqnarray}
E_G &=& -\frac{3}{5} \frac{G\,m^2}{R},
\end{eqnarray}
where,
G is the gravitational constant.
The computed values for the respective cores are given in
Table\,\ref{table_CoreSummary}.

For comparison, we also calculate the parameters V$_A$, M$_A$, KE terms
(E$_{KE}$, E$_{KE,T}$, and E$_{KE,NT}$), and E$_B$ for the respective filament
neighbourhoods of the cores. For the filamentary part, we estimate the
volume number density \nHtwo\, by dividing the mean column density of the
neighbourhood by the width of the filament,
i.e. $\nHtwo$=$\NHtwo/W_{fil}$, where W$_{fil}$=$0.1\,pc$. The
mass of the neighbourhood was estimated using :
$m$=$\muHtwo \mathrm{m_{H}} Area_{\mathrm{pixel}} \Sigma \NHtwo$,
where Area$_{\mathrm{pixel}}$ is the area of each pixel and
$\Sigma \NHtwo$ is the sum of column densities of the constituent pixels.
The results are listed in Table\,\ref{table_CoreSummary}.

\begin{figure}
%\centering
\hspace{-25pt}
\includegraphics[width=1.1\columnwidth]{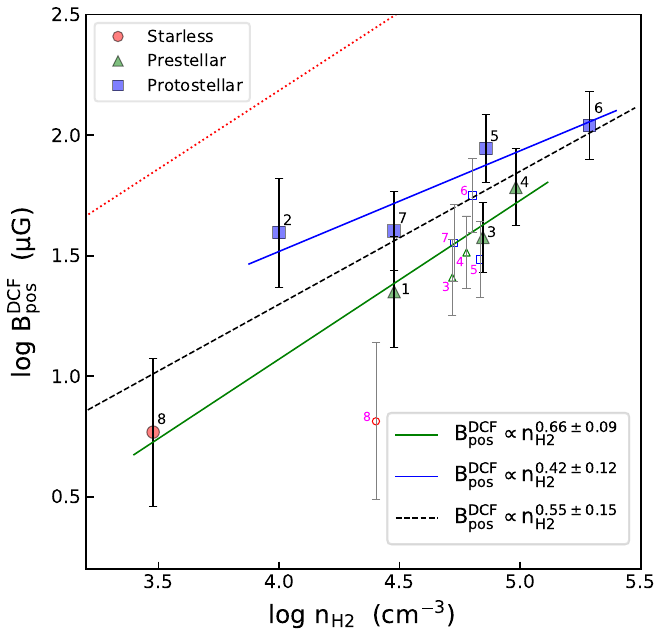}
\caption{
Log-log relation between \Bpos\, and n$_{\scriptscriptstyle \mathrm{H2}}$ for
the HGBS cores.
The red dotted line is the relation from
\citet[][slope$\approx$0.65 derived from Zeeman measurements]{Crutcher_2010ApJ}.
The filled symbols denote the core values and unfilled symbols the
respective neighbourhood values. Both are labelled.
The green line, blue line, and the black dashed line show the best straight
line fit between log\,\BposDCF\, and log\,n$_{\scriptscriptstyle \mathrm{H2}}$
for the prestellar/starless cores (1, 3, 4, and 8),
protostellar cores (2, 5, 6, and 7) and all the cores together, respectively.
}
\label{fig_Bn_Cores}
\end{figure}

\subsection{B-n Relation}
\label{section_B_n_Relation_cores}

Figure \ref{fig_Bn_Cores} shows the relation between \BposDCF\, and \nHtwo\,
on a log-log plot.
The dotted line is the relation from
\citet[][derived from Zeeman measurements]{Crutcher_2010ApJ} (using
\nHtwo=n$_{\scriptscriptstyle \mathrm{H}}$/2). A straight line fit gives the
power law of the form \Bpos$\propto$\,\nHtwo$^{\!\!\!\mathrm{\kappa}}$.
A value of $\kappa$=2/3 is predicted for the case of spherical collapse with
flux freezing \citep{Mestel_1966MNRAS}, i.e. the weak field case;
while $\kappa<0.5$ is predicted for the strong field case with dominance of
ambipolar diffusion \citep{Mouschovias_1999ASIC}.
We carried out the fit (the error in \BposDCF\, was taken into account)
for the
prestellar/starless cores (cores 1, 3, 4, and 8),
protostellar cores (cores 2, 5, 6, and 7), and
all the cores together to obtain $\kappa$ values of
0.66$\pm$0.09, 0.42$\pm$0.12, and 0.55$\pm$0.15, respectively
(see Section \ref{section_discussion_Bn} for a discussion of the possible
physical interpretation).

The relation is affected by uncertainties in the magnetic field strength
as well as the volume density. According to \citet{Konyves_2020AA_HGBS}, the
HGBS core mass (and thus the volume density) estimates are affected by the
following systematic effects : uncertainty in dust opacity law leading to
mass uncertainty factor of $\sim$1.5-2, and use of line-of-sight averaged
dust temperatures.
As for the magnetic field estimates, Equation \ref{eqn_Bpos} for DCF assumes
a \textquotedblleft correction factor\textquotedblright\, of
0.5 \citep{Crutcher_2004ApJ}, relying on the simulations of
\citet{Ostriker_2001ApJ}.
However, studies have shown that in dense self-gravitating cores, the
correction factor can decrease upto $\sim$\,0.28
\citep{Kwon_Bistro_OphA_2018ApJ,Liu_NumSim_2021ApJ,Liu_2022FrASS,Myers_DCFEstimates_2024ApJ}.
Another source of uncertainty could be that the assumption of equipartition
of turbulent kinetic energy and turbulent magnetic energy might not hold true
-- at least for the super-Alfvenic cores in our sample
\citep{FalcetaGoncalves_2008ApJ,Liu_NumSim_2021ApJ} -- leading to an
overestimated field strength for these cores using DCF method.

%%%%%%%%%%%%%%%%%%%%%%%%%%%%%%%%%%%%%%%%%%%%%%%%%%%%%%%%%%%%%%%%%%%%%%%%%%%%%%%%

\section{Discussion}
\label{section_discussion}

Our analysis of the region reveals the complex nature of magnetic fields which
thread this filamentary cloud region, and affect its dynamics as well as of the
cores embedded therein.

\subsection{The fragmenting filament}
\label{section_discussion_fragmentingfilament}

The approximately 1.25\,pc long filamentary structure displays a line mass
($\sim$80\,\msun\,pc$^{-1}$) much larger than the critical magnetic line mass
($\sim$37\,\msun\,pc$^{-1}$) and is thus expected to be in a fragmentation
and collapse stage.
The presence of cores at different evolutionary stages probably attest to this
fact.

The magnetic field half-vectors from \emph{Planck} show their orientation
to be perpendicular to the filament's overall longitudinal direction
(see Figures \ref{fig_StokesI_FullFoV} and \ref{fig_quiverPlot_PF}). This is
in line with observations in literature wherein \Bpos\, has been mostly found
perpendicular to high-density supercritical filaments
\citep{Pineda_PPVII_2023ASPC,Planck_XXXII_2016AA}.
Albeit at low-resolution, the vectors do not show any
reorientation of \Bpos\, from perpendicular to parallel
\citep[\textquoteleft U-shaped\textquoteright\, morphology,][]{Gomez_2018MNRAS}
as density increases at the filament crest.
Even the JCMT vectors give no such indication.
Three out of four Cores in the dense central part of the filament -- namely
cores 3, 4, and 6 -- show their constituent pixels' JCMT vectors
(Figure \ref{fig_quiverPlot_PF}) to be mostly aligned with the \emph{Planck}
vectors, i.e. perpendicular to the filament's overall longitudinal direction.
This could indicate no longitudinal flow of material along
the filament to its \textquotedblleft high-density\textquotedblright\, part.
We add the caveat that the vectors in core 5 could be partially aligned with
the filament crest orientation locally
(see Figures \ref{fig_CrestLine}, and \ref{fig_CrestVar_BpaMinusCrestAngle}).
The overall magnetic field orientation of the cores with respect to the
filament crest does not appear to show any particular trend
(see Figure \ref{fig_CrestVar_BpaMinusCrestAngle}).
As such, it seems to agree with simulations which suggest
random orientation, positing that the randomness of the magnetic fields in
the cores is inherited from the transonic velocity fluctuations of their
host magnetized filaments during the fragmentation
\citep{Misugi_2023ApJ,Misugi_2024ApJ}.

According to evolutionary stage, the filamentary cloud consists of
1 starless core, 3 prestellar cores, and 4 protostellar cores.
Except for core 1, the rest (cores 2 to 8) lie along the contiguous
filamentary structure.
The prestellar cores 3 and 4, and the starless core 8 appear to have their
major axis aligned along the filament longitudinal axis. This is also borne
out by simulations wherein the cores formed along a fragmenting filament have
ellipsoidal shapes with longer axis aligned with filament axis
\citep{Misugi_2023ApJ}.
Protostellar core 5 has no 70\micron\, detection according to the HGBS catalog,
unlike other protostellar cores; yet is associated with an outflow according
to ALMA observations. Hence it is likely the youngest among protostellar cores
in this filamentary region. Core 5 is also the location of highest column
density in the region, albeit only marginally larger than that for
(protostellar) core 6
\citep[identified as a hot corino in literature;][]{Liu_ALMASOP_2025ApJ}.
For core 6, the outflow
\citep[from][position angle = 149.1\,deg]{Liu_ALMASOP_2025ApJ}
does not appear to be aligned with either the orientation of the core
($\sim$16\,deg) or \Bpos\,($\sim$18$\pm$2\,deg)
(see Figure \ref{fig_CoreNbrhoodCompare} and Table \ref{table_extCoreSummary}).

The starless and prestellar cores -- i.e. cores 1, 3, 4, and 8 -- have larger
angular dispersions than protostellar cores ($\delta\theta_{\mathrm{extCore}}$
in Table \ref{table_extCoreSummary}). This is likely so as the field lines are
too weak to resist the effect of turbulence in younger stages, and as the
gravitational collapse proceeds, the field lines become more ordered due to
the dissipation of turbulence \citep{Crutcher_2012ARAA}. This can be clearly
seen on a visual examination of the field lines as well
(see Figure \ref{fig_CoreNbrhoodCompare}).

\subsection{Depolarisation in the cores}
\label{section_discussion_depolarisation}

Examination of polarisation parameters for the filament and the core regions
shows a smaller polarisation fraction for the cores. The polarisation fraction
as a function of intensity is given by $\PF \propto I^{-\alpha}$, wherein we
calculated the value of $\alpha$ for the
filament (0.71$\pm$0.08) and the cores (0.85$\pm$0.10)
separately by applying the Ricean Mean Model (Section \ref{section_PFvsI}).
The $\alpha$ values, along with the smaller polarisation fraction of the
cores, suggest a combination of loss of grain alignment and change in
structure of the field between the filament and the cores.

Looking at only the core values, we find that the polarisation
parameters for the protostellar cores 5, 6, and 7 are similar to that
of their respective neighbourhoods, in contrast to prestellar cores
(Figure \ref{fig_PFDiffVsBpaDiff}).
This suggests that among the possible causes of depolarisation
-- namely, loss in grain alignment, dust grain destruction in the vicinity
of protostars, and integration over field geometry along the LOS
\citep{Pattle_PPVII_2023ASPC} --
the latter might not be the cause for
depolarisation in the protostellar cores.
For this, we refer to the
\textquotedblleft two-component model along the LOS\textquotedblright\,
from \citet[see their Appendix A and Figure A.2]{Planck_XXXIII_2016AA}.
According to this model, the depolarisation factor due to the rotation of
\Bpos\, is $\sim$0.9-1 for a small angle difference from the background.
Hence it is reasonable to conclude that the depolarisation for the
protostellar cores (5, 6, and 7) is likely due to the loss in grain alignment
and/or dust grain destruction in the vicinity of these protostars.

\subsection{Transition from weak to strong field with core evolution ?}
\label{section_discussion_Bn}

The B-n relation for the cores shows a clear distinction
between the power law indexes for the prestellar/starless cores and the
protostellar cores (Figure \ref{fig_Bn_Cores}).
The index for the prestellar/starless cores
($\kappa$\,$\sim$0.66$\pm$0.09)
indicates a weak field case.
The value is consistent with that found by \citet{Crutcher_2010ApJ} using
Zeeman
observations ($\kappa\approx0.65$), as well as the index calculated by
\citet{Myers_DenseCores_2021ApJ} (0.66$\pm$0.05) using a literature-compiled
sample of 17 dense cores (using DCF method).
Though it should be noted that \citet{Myers_DenseCores_2021ApJ} conclude from
their
analysis that a weak field is not a requisite for their calculated relation,
and the relation can be statistically explained on the basis of distribution
of mass and volume density for a sample of approximately magnetically
critical cores which are strongly bound and spheroidal with a large range
of radii.

On the other hand, the more evolved protostellar cores
($\kappa$\,$\sim$0.42$\pm$0.12)
appear to be consistent with the strong field case.
The overall value ($\kappa$\,$\sim$0.55$\pm$0.15) lies between the two limits,
and is very similar to the index of 0.57 found by \citet{Liu_DCF_2022ApJ}
using (recalculated) previous polarised dust observations of a sample at
clump and core scales (using DCF).

This analysis indicates a possible scenario wherein the prestellar/starless
core stage could be dominated by weak field case, and there is a
\textquoteleft switch\textquoteright\, to strong field domain as the core
moves along its evolutionary path to protostellar stage.
Supporting evidence for this is also
provided by our mass to flux ratio ($\lambda$) values
(Table \ref{table_CoreSummary}).
According to \citet{Crutcher_2012ARAA}, $\lambda$ for the cores in the
strong field case is $\sim$1 or only slightly $>$1.
But for the weak field case, the region is supercritical, and $\lambda$ may
take any value $>$1. That we see $\lambda^{\textsc{dcf}}_{obs}$ for
prestellar cores larger than the protostellar cores on an average in
Table \ref{table_CoreSummary} could be a reflection of this.
Moreover, if corrections for total magnetic field and orientation are applied
(Section \ref{section_MassToFluxRatio}), only the prestellar cores (3 and 4)
appear to be critical and rest subcritical.
It is also noteworthy that the magnetic field strengths for the
prestellar/starless cores here are also weaker than those for the
protostellar cores.

Hence it could be that as a (prestellar) core permeated by a weak magnetic
field collapses with flux-conservation \citep{Mestel_1966MNRAS}, then at some
stage during the collapse, the (cross-sectional) area will become small but
the total flux being the same, the field strength associated will now be
higher. At this stage, the prestellar core is probably dense enough to
transition to a protostellar stage, and the field strong enough to be called
strong field case.

Another possible reason for the different slope of the protostellar
cores and the prestellar/starless cores, as well as for the larger
mass to flux ratio of the prestellar cores could be a difference of condition
of the respective local regions. The eastern part of the filament
(see Figure \ref{fig_StokesI_FullFoV}) which harbours the prestellar
cores 3 and 4 (see Figure \ref{fig_quiverPlot_PF}) appears to to be
joined by a secondary filament from its north. The flow of material along
the secondary filament to the eastern part of the main filament could be
affecting the physical conditions there. The western part with protostellar
cores 5, 6, and 7 is devoid of such an effect.

\subsection{Limitations of our Dust Polarisation Analyses}
\label{section_limitations}

This study is mainly limited by low sensitivity in many parts of the filament,
leaving pixel gaps for some of the cores and their neighbourhood regions,
resulting in poor statistics.
Moreover SCUBA-2/POL-2 DAISY mapping is only reliable for the central region
of the field of observation, and thus one cannot take a complete
view of the larger region which shows secondary filaments joining the main
filamentary structure (see Figure \ref{fig_StokesI_FullFoV}) in this study.

Furthermore, calculation of magnetic field strengths using the DCF method
(and its variants) assumes turbulence to be the reason for distortion of the
field over its mean background value. It discounts complicated structures
which can arise in evolved cores due to phenomena such as outflows and
gravitational collapse (like hourglass morphology)
(also see Section \ref{section_DCFIssues}). Lastly, the magnetic
field strength which is used in this work is the plane-of-sky
component. Our study thus lacks any information about three dimensional
structure of the magnetic field.
This leads to incomplete information about the measured properties
in some instances, such as
serendipitous orientation of the field vector exactly along the line-of-sight
wherein no polarisation would be observed;
and averaging out of field vectors along the line-of-sight, to mention a few.

%%%%%%%%%%%%%%%%%%%%%%%%%%%%%%%%%%%%%%%%%%%%%%%%%%%%%%%%%%%%%%%%%%%%%%%%%%%%%%%%

\section{Summary and Conclusions}
\label{section_summaryconclusions}

In this paper, we have carried out an analysis of polarised dust emission
at 850\micron\, from a filament
(also associated with \object{PGCC G203.21-11.20})
in the northern part of the Orion\,B molecular cloud complex.
The region was observed from JCMT SCUBA-2/POL-2 polarimeter.
The main findings can be summarised as follows :
\begin{enumerate}
\item
The line mass of the nearly 1.25\,pc long filamentary structure is calculated
to be $\sim$80\,\msun\,pc$^{-1}$, which is larger than the critical line mass
of $\sim$37\,\msun\,pc$^{-1}$ for the magnetized filament
(\Bpos$\sim$31\,$\text{\textmu}$G),
thus making it unstable to gravitational collapse.
The filament is found to be fragmented into 1 starless, 3 prestellar, and
4 protostellar cores.
%identified from the HGBS catalog and outflow literature from ALMA.
\item
The total 198 detected pixels at our SNR criteria were classified into 125
filament and 73 core pixels. The mean \bpa\, for the filament and the
cores was calculated to be 50$\pm$3\,degrees and 30$\pm$4\,degrees,
respectively; while \PFdeb\, was found to be $\sim$5.3$\pm$0.3\% and
$\sim$3.2$\pm$0.3\%, respectively, suggesting that they represent distinct
physical conditions.
\item
The polarisation fraction of the cores was found to be correlated with their
location rather than their type. Cores (3 to 6) in the denser part
(\NHtwo$\gtrsim$2$\times$10$^{22}$cm$^{-2}$) have a nearly constant
\PFdeb\, of $\sim$1$^{+0.7}_{-0.1}$\%, while those at the filament edges
vary in the range $\sim$6-11\%.
\item
Protostellar cores were found to have their magnetic field orientation
more aligned with their respective neighbourhood filament's
magnetic field orientation
than prestellar cores in this region.
\item
The Ricean Mean Model fit to the relation PF$\propto$I$^{-\alpha}$ returns
$\alpha$=0.71$\pm$0.08 and 0.85$\pm$0.10 for the filament and the cores,
respectively.
This result, along with the smaller polarisation fraction of the cores,
suggests a combination of loss of grain alignment and change in structure
of the field between the filament and the cores.
\item
The prestellar cores and the (sole) starless core were found to have larger
angular dispersion than the protostellar cores.
Using DCF,
the plane-of-sky field strength (\BposDCF) of
the starless core was calculated to be $\sim$6\,$\text{\textmu}$G;
the prestellar cores to be in the range $\sim$22-61\,$\text{\textmu}$G; and
the protostellar cores to be in the range $\sim$39-110\,$\text{\textmu}$G.
Thus, protostellar cores have stronger field strength than prestellar cores.
Protostellar cores 5 and 6 have the highest \BposDCF\, of
87.9$\pm$28.4\,$\text{\textmu}$G and 109.8$\pm$36.0\,$\text{\textmu}$G,
respectively.
%ALMA literature posits both to be associated with outflows
%(G203.21-11.20W1 and G203.21-11.20W2 corresponding to core 5 and 6,
%respectively; with G203.21-11.20W2 (core 6) classified as a hot corino).
\item
The mass-to-flux ratio using \BposDCF\, as the field strength yields all cores,
except core 2 and 8, to be critical. However, using the total field strength
$|$\textbf{B}$|$ (i.e. $|$\textbf{B}$|$=($4/\pi$)\Bpos) and the statistical
correction for the orientation (i.e. $\lambda=\lambda_{obs}/3$), all except
the prestellar cores 3 and 4 are found to be subcritical.
%Cores 3 and 4 are also the most massive cores with masses (from HGBS) of
%8.91$\pm$1.80\,\msun\, and 9.13$\pm$2.02\,\msun, respectively.
They are also the only cores which are super-Alfvenic (M$_\mathrm{A}>1$)
in our dataset.
\item
The B-n relation (\Bpos$\propto$\,\nHtwo$^{\!\!\!\mathrm{\kappa}}$) for the
prestellar/starless cores, protostellar cores, and
all the cores together returns a power law index ($\kappa$) of
0.66$\pm$0.09, 0.42$\pm$0.12, and 0.55$\pm$0.15, respectively.
The fitting suggests a transition from weak field case to strong field case
as the cores evolve from the prestellar to protostellar phase.
\end{enumerate}

The study of magnetic field in a single fragmenting filament populated by
cores at various evolutionary stages has provided us important insights on
the intertwined nature of (prestellar to protostellar) core evolution and
(weak to strong) field transition.
To better understand the fragmentation of magnetized filaments and subsequent
evolution of cores, prospective work would not only require polarisation
mapping of the larger FoV, but also
at a resolution which allows one to study structures and phenomena
associated with protostellar objects, such as hourglass morphology and
effect of outflows.
Furthermore, core-wise analysis of magnetic fields could be extended to other
filamentary molecular clouds for sampling different environments.
The role of current facilities like ALMA;
upcoming submillimeter facilities like the
Large Submillimeter Telescope (LST) and
Atacama Large Aperture Submillimeter Telescope (AtLAST);
and instruments like the PRIMA Polarimetric Imager and
the next generation JCMT/SCUBA-3
would be crucial in future polarisation studies of star forming regions.

%%%%%%%%%%%%%%%%%%%%%%%%%%%%%%%%%%%%%%%%%%%%%%%%%%%%%%%%%%%%%%%%%%%%%%%%%%%%%%%%

\begin{acknowledgments}
We thank the anonymous referee for providing suggestions which helped in
improving the paper.
This research is partially supported by the NAOJ University Support Expenses
(FY2025).
Data analysis was in part carried out on the Multi-wavelength Data Analysis
System operated by the Astronomy Data Center (ADC), National Astronomical
Observatory of Japan.
The James Clerk Maxwell Telescope is operated by the East Asian Observatory
on behalf of
the National Astronomical Observatory of Japan;
Academia Sinica Institute of Astronomy and Astrophysics;
the Korea Astronomy and Space Science Institute;
the National Astronomical Research Institute of Thailand;
Center for Astronomical Mega-Science (as well as the National Key R\&D Program
of China with No.\,2017YFA0402700).
Additional funding support is provided by the Science and Technology Facilities
Council of the United Kingdom and participating universities and organizations
in the United Kingdom, Canada, and Ireland.
Additional funds for the construction of SCUBA-2 were provided by the Canada
Foundation for Innovation.
The authors wish to recognize and acknowledge the very significant cultural
role and reverence that the summit of Mauna Kea has always had within the
indigenous Hawaiian community. We are most fortunate to have the opportunity
to conduct observations from this mountain.
\end{acknowledgments}

\facilities{JCMT, Herschel, IRSA}

%%%%%%%%%%%%%%%%%%%%%%%%%%%%%%%%%%%%%%%%%%%%%%%%%%%%%%%%%%%%%%%%%%%%%%%%%%%%%%%%
\clearpage
\appendix

%%%%%%%%%%%%%%%%%%%%%%%%%%%%%%%%%%%%%%%%

\section{Calculation of Planck polarisation Angles}
\label{appendix_Planck}

The stokes Q and U images from \emph{Planck} are in Galactic coordinates, have
a beamsize of \textsc{fwhm}$\sim$5\arcmin, and follow the
\textquotedblleft COSMO\textquotedblright\, convention \citep{Planck_I_2016AA}
rather than the IAU convention followed by JCMT. Hence, the polarisation angle
($\Theta$) in IAU convention from the \emph{Planck} stokes parameters is given
by \citep{Planck_XIX_2015AA} :
\begin{equation}
\label{eqn_pa_Planck}
\PA = \mathrm{\frac{1}{2}~tan^{-1}} \left( \frac{-U}{Q} \right),
\end{equation}
where \PA\, is measured from the north Galactic pole, and increases
counterclockwise towards Galactic east. The orientation of the magnetic
field vectors, as for JCMT, is given by $\bpa=\PA+\pi/2$
(Equation \ref{eqn_bpa}).
We extract the \emph{Planck} half-vectors in the Galactic coordinate system
as ds9 \citep{ds9_2000ascl} regions, and use them for visual examination of
large scale magnetic field in Section \ref{section_polarisationMorphology}
and Figure \ref{fig_quiverPlot_PF}.

%%%%%%%%%%%%%%%%%%%%%%%%%%%%%%%%%%%%%%%%

\section{Comparison of different SNR limits}
\label{appendix_compareSNRlimits}

\begin{figure*}[b]
\centering
%
%Section 3.2 from https://cs.brown.edu/about/system/managed/latex/doc/subfigure.pdf
\renewcommand{\thesubfigure}{(a)}
\subfigure
{
%\centering
\includegraphics[width=0.47\columnwidth]{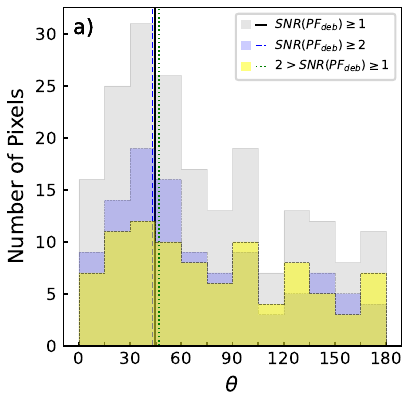}
\label{fig_histogram_bySNR_Bpa}
}
\renewcommand{\thesubfigure}{(b)}
\subfigure
{
%\centering
\includegraphics[width=0.47\columnwidth]{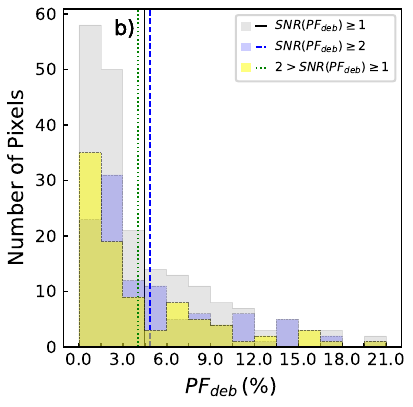}
\label{fig_histogram_bySNR_PFdeb}
}
\caption{
Distributions by different SNR criteria of \PFdeb. The vertical lines mark the
respective means for different criteria.
\emph{(a)} Distribution of magnetic field orientations (\bpa).
\emph{(b)} Distribution of \PFdeb.
}
\label{fig_histogram_bySNR}
\end{figure*}

Figure \ref{fig_histogram_bySNR} shows the histogram of \bpa\, and \PFdeb\, for
different SNR criteria of \PFdeb. It should be noted that all the data points
used for the histogram analysis here already satisfy the criteria
\PFdeb$\leq$20\% and SNR(I)$\geq$9. In this section, we only examine the
distribution of data points satisfying different SNR criteria on \PFdeb.
For \bpa\, in Figure \ref{fig_histogram_bySNR_Bpa}
(\PFdeb\, in Figure \ref{fig_histogram_bySNR_PFdeb}),
the mean angles (polarisation fractions) were found to be
$\sim$47$\pm$6\,degrees (4.1$\pm$0.45\%) and
$\sim$43$\pm$2\,degrees (4.9$\pm$0.23\%) for
2$>$SNR(\PFdeb)$\geq$1 and SNR(\PFdeb)$\geq$2, respectively.
The mean value taking all the data points into consideration,
i.e. SNR(\PFdeb)$\geq$1, is calculated to be
$\sim$45$\pm$2\,degrees for \bpa\, (4.5$\pm$0.24\% for \PFdeb).
The mean values for the different SNR(\PFdeb) criteria are thus found to be
within one binwidth -- 15\textdegree\, for \bpa\, and 1.5\% for \PFdeb. Moreover,
the histograms display a similar shape for the three SNR criteria.
For \bpa, the maxima lies in the 30-45\,degree\, bin, and the second highest
peak in the 90-105\,degree bin. The histogram for \PFdeb\, shows that most of
the pixels lie in the 0-3\% bins, followed by a rapid decrease of pixels in the
subsequent bins.
The above analysis demonstrates that pixels with low SNR,
i.e. 2$>$SNR(\PFdeb)$\geq$1, appear to be qualitatively no different from
those with higher SNR.
Hence, we use a \PFdeb\, SNR limit of $\geq$1 for selection of pixels (over and
above the criteria of \PFdeb$\leq$20\% and SNR(I)$\geq$9).

%%%%%%%%%%%%%%%%%%%%%%%%%%%%%%%%%%%%%%%%

\begin{figure*}
\centering
%
%Section 3.2 from https://cs.brown.edu/about/system/managed/latex/doc/subfigure.pdf
\renewcommand{\thesubfigure}{(top)}
\subfigure
{
%\centering
\includegraphics[width=\textwidth]{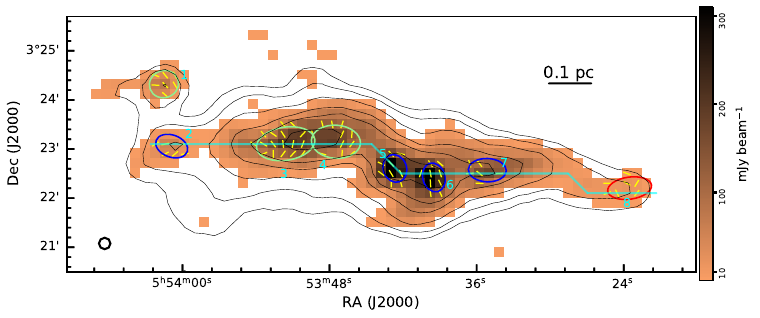}
\label{fig_CrestLine}
}
\renewcommand{\thesubfigure}{(bottom)}
\subfigure
{
%\centering
\includegraphics[width=\textwidth]{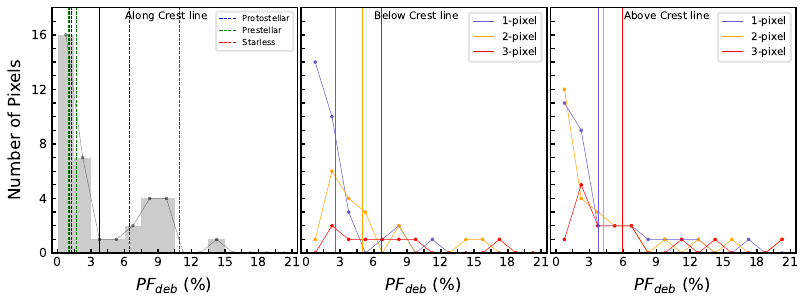}
\label{fig_radialVar_PF}
}
\caption{
\emph{(top)}
Stokes I image at SNR$\geq$5. Red contours mark column density values at
0.55, 0.7, 1, 1.5, 2, and 3$\times$10$^{22}$cm$^{-2}$. Red, green, and blue
ellipses are the starless, prestellar, and protostellar cores, respectively.
\bpa\, of the core pixels have been shown by yellow line segments. Cyan line
marks the crest of the filament.
Black circle on bottom left shows \emph{JCMT} beam ($\sim$14\arcsec).
\emph{(bottom)}
Histogram showing the distribution of \PFdeb\, of the pixels along the crest
(left), below the crest (middle), and above the crest (right). In the left
panel, the dashed lines mark the mean \PFdeb\, of the cores along the crest
line (i.e. cores 2 to 8). Black vertical line marks the mean \PFdeb\, of all
the crest pixels on the left panel. The mean of \PFdeb\, of pixels which are
1-, 2-, and 3-pixel(s) away from the crest is shown by the respective colored
line in the middle and right panel.
}
\label{fig_CrestLine_RadialPF}
\end{figure*}
%\clearpage

\begin{figure*}
\centering
%
%Section 3.2 from https://cs.brown.edu/about/system/managed/latex/doc/subfigure.pdf
\renewcommand{\thesubfigure}{(left)}
\subfigure
{
%\centering
\includegraphics[width=0.47\columnwidth]{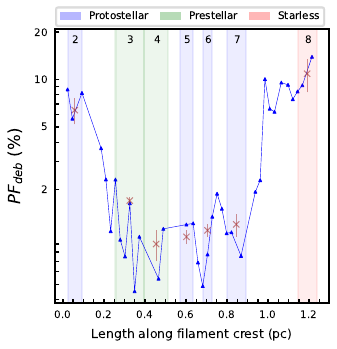}
\label{fig_CrestVar_PF}
}
\renewcommand{\thesubfigure}{(right)}
\subfigure
{
%\centering
\includegraphics[width=0.47\columnwidth]{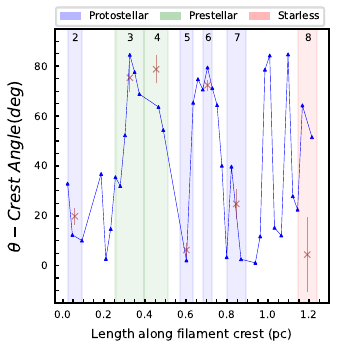}
\label{fig_CrestVar_BpaMinusCrestAngle}
}
\caption{
Variation of
\emph{(left)}
\PFdeb\, and \emph{(right)} \bpa-Crest\,Angle along the crest of the filament
shown in Figure \ref{fig_CrestLine}.
Color spans mark the span of the cores as they are encountered while traversing
along the crest (i.e. cores 2 to 8).
The mean quantity for the respective cores has been marked as crosses.
}
\label{fig_CrestVar}
\end{figure*}

\section{Variation from Filament Crest}
\label{appendix_radialVariation}

In this section we examine how the polarisation parameters vary as one traverses
along the filament crest and radially from the filament crest.
Figure \ref{fig_CrestLine} shows the crest of the filament overlaid on the
stokes I image. The background stokes I image shows only those pixels which
have SNR(I)$\geq$5. The crest was traced using a combination of stokes I
image and the column density contours.
The extended-cores and the magnetic field segments of the respective
extended-cores have also been shown on the image. Red, green, and blue
ellipses denote the starless, prestellar, and protostellar cores,
respectively.
Except core 1, all the cores (from 2-8) lie along the crest.
Note that not all pixels along the filament crest or in its surroundings had
detections at our requisite criteria (see Figure \ref{fig_quiverPlot_PF} for
the pixels which are used in the analyses).

In Figure \ref{fig_CrestLine_RadialPF}, we look at the radial variation
of the \PFdeb\, from the crest. For this, we examine the variation of \PFdeb\,
for
the pixels constituting the crest; for the pixels lying 1-, 2-, and 3-pixels
away from the crest towards the the south (\textquoteleft below crest
line\textquoteright); and for the pixels lying 1-, 2-, and 3-pixels away from
the crest towards the the north (\textquoteleft above crest
line\textquoteright).
The left panel in Figure \ref{fig_CrestLine_RadialPF} shows the distribution of
pixels on the filament crest. The black vertical line marks the mean.
The middle panel shows the distribution of the pixels below the crest which are
radially 1-, 2-, and 3-pixel away. The mean of each distribution is also
marked. Similarly, the radial variation above the crest line is shown in the
right panel. It can be seen that as one moves radially away from the crest, the
polarisation fraction increases. This is true both below and above the crest
line.

\subsection{Traversal along the filament crest}
\label{appendix_radialVariation_TraversalAlong}

We also examine the variation of \PFdeb\, as one traverses along the crest
(from east to west). The same is shown in Figure \ref{fig_CrestVar} as a
function of projected distance in pc. The color spans here show the span of the
cores which one encounters (2-8, in order) while traversing.
For comparison, the mean \PFdeb\, for the individual cores have been marked as
crosses. A trend can clearly be seen wherein the polarisation fraction is
highest at the edges of the filament and shows a decline as one moves towards
the central part. Thus, both in Figure \ref{fig_CrestLine_RadialPF} and
\ref{fig_CrestVar_PF}, it is apparent that the lowest polarisation fraction
coincides with the densest part of the filament and vice-versa.

Another quantity which is of interest is the orientation of magnetic field
segments with respect to the crest of the filament. This relative orientation
is plotted in Figure \ref{fig_CrestVar_BpaMinusCrestAngle}. While overall the
plot seems to show wide fluctuation in the relative orientation, certain trends
can be observed for individual cores. For example, cores 3, 4, and 6 are nearly
perpendicular to the filament crest, while the rest show smaller deviation in
the range $\sim$0-20\,degree. The crest of core 3 shows the magnetic field
segments rotating from nearly parallel to the crest (also see Figure
\ref{fig_quiverPlot_PF}) to perpendicular to the crest at approximately the
middle of the filament and then seemingly rotating back to parallel
orientation. Since otherwise cores 3, 4, 5, and 6 have similar properties when
it comes to column density, \PFdeb, and \bpa\, (see Figures \ref{fig_PFVsBpa}
and
Table \ref{table_extCoreSummary}), the main reason for the different
relative orientation of core 5 from its neighbours in Figure
\ref{fig_CrestVar_BpaMinusCrestAngle} seems to be the orientation of the
filament in the local region. Hence, at least for these four cores (3 to 6) in
the dense filamentary region, the orientation of the filament does not appear
to play a part in influencing their properties.

%%%%%%%%%%%%%%%%%%%%%%%%%%%%%%%%%%%%%%%%

\section{Calculation of standard errors}
\label{appendix_Errors}

\subsection{Error on Circular mean}
\label{appendix_Errors_CircMean}

If we have \emph{n} pixels, and the polarisation angle associated with the
pixels is $\Theta_i=(\Theta_1, \Theta_2, ..., \Theta_n)$, then the mean
polarisation direction is given by :
\begin{equation}
\label{eqn_Theta_mean}
\Theta_{mean} = \frac{1}{2} arctan\left( \frac{\bar{\textsc{s}}}{\bar{\textsc{c}}} \right),
\end{equation}
where,
\begin{eqnarray}
\bar{S} &=& \frac{1}{n} \Sigma_{i=1}^{n} sin(2 \Theta_i), and \\
\bar{C} &=& \frac{1}{n} \Sigma_{i=1}^{n} cos(2 \Theta_i)
\end{eqnarray}
For error on the circular mean ($\Theta_{mean}$) of the polarisation angles
($\Theta_i$), we can use the definitions of the stokes parameters.
For each pixel \emph{i} (taking \PF$_i$ as the polarisation fraction),
\begin{eqnarray}
U_i = \PF_i I_i\,sin\,2\Theta_i = \sqrt{U_i^2+Q_i^2}\,sin\,2\Theta_i, &and&\,\, Q_i = \PF_i I_i\,cos\,2\Theta_i = \sqrt{U_i^2+Q_i^2}\,cos\,2\Theta_i, \\
\Rightarrow U_i/\sqrt{U_i^2+Q_i^2} = sin\,2\Theta_i, &and&\,\, Q_i/\sqrt{U_i^2+Q_i^2} = cos\,2\Theta_i
\end{eqnarray}
It can be seen that :
\begin{eqnarray}
\bar{S} &=& \frac{1}{n} \Sigma_{i=1}^{n} U_i/\sqrt{U_i^2+Q_i^2} = \frac{1}{n} \Sigma_{i=1}^{n} u_i, and \\
\bar{C} &=& \frac{1}{n} \Sigma_{i=1}^{n} Q_i/\sqrt{U_i^2+Q_i^2} = \frac{1}{n} \Sigma_{i=1}^{n} q_i
\end{eqnarray}
Since we are provided with variance ($\sigma_U^2$ and $\sigma_Q^2$) images for
U and Q, we can calculate the error on u and q ($\sigma_u$ and $\sigma_q$)
for each \emph{i-th} pixel by standard error propagation :
\begin{eqnarray}
\sigma_{u_i}^2 &=& \left( \frac{\partial u_i}{\partial U_i} \right)^2 \sigma_{U_i}^2 + \left( \frac{\partial u_i}{\partial Q_i} \right)^2 \sigma_{Q_i}^2, and \\
\sigma_{q_i}^2 &=& \left( \frac{\partial q_i}{\partial U_i} \right)^2 \sigma_{U_i}^2 + \left( \frac{\partial q_i}{\partial Q_i} \right)^2 \sigma_{Q_i}^2 .
\end{eqnarray}
Error propagation on the mean of u$_i$ and q$_i$ will then give error on
$\bar{S}$ and $\bar{C}$ as :
\begin{eqnarray}
\sigma_{\bar{\textsc{s}}}^2 &=& \displaystyle\sum_{i=1}^{n} \sigma_{u_i}^2/n^2, and\,\, \sigma_{\bar{\textsc{c}}}^2 = \displaystyle\sum_{i=1}^{n} \sigma_{q_i}^2/n^2.
\end{eqnarray}
Thus, once we have the values of $\bar{S}$, $\bar{C}$,
$\sigma_{\bar{\textsc{s}}}$, and $\sigma_{\bar{\textsc{c}}}$,
then the error on $\Theta_{mean}$ is given by (similar to
equation\,\ref{eqn_ErrorBpa}) :
\begin{eqnarray}
\sigma_{_{\Theta_{mean}}} = \frac{1}{2} \frac{\sqrt{\bar{\textsc{s}}^2 \sigma_{\bar{\textsc{c}}}^2 + \bar{\textsc{c}}^2 \sigma_{\bar{\textsc{s}}}^2}}{\bar{\textsc{s}}^2 + \bar{\textsc{c}}^2}
\end{eqnarray}

\subsection{Error on \Bpos\,}
\label{appendix_Errors_Bpos}

B$_{\mathrm{pos}}^{\textsc{dcf}}$ and B$_{\mathrm{pos}}^{\textsc{st}}$,
given by
Equation \ref{eqn_Bpos} and \ref{eqn_Bpos_ST}, respectively, can be expanded
as follows for our case :
\begin{eqnarray}
\label{eqn_Bpos_Expanded}
B_{\mathrm{pos}}^{\textsc{dcf}} &\approx& 9.3\, \sqrt{\frac{\NHtwo}{W_{fil}}}~\frac{\Delta V_{NT}}{\delta \theta}, and \\
B_{\mathrm{pos}}^{\textsc{st}}  &\approx& 1.76\,\sqrt{\frac{\NHtwo}{W_{fil}}}~\frac{\Delta V_{NT}}{\sqrt{\delta \theta}}
\end{eqnarray}
Error propagation on the above gives :
\begin{eqnarray}
\sigma_{\scriptscriptstyle B_{pos,\textsc{dcf}}} &=& B_{\mathrm{pos}}^{\textsc{dcf}} \sqrt{\frac{\sigma_{\NHtwo}^2}{4\NHtwo^2}+\frac{\sigma_{\scriptscriptstyle W_{fil}}^2}{4 W_{fil}^2}+\frac{\sigma_{\scriptscriptstyle \Delta V_{NT}}^2}{\Delta V_{NT}^2}+\frac{\sigma_{\scriptscriptstyle \delta\theta}^2}{\delta\theta^2}},\,and \\
\sigma_{\scriptscriptstyle B_{pos,\textsc{st}}}  &=& B_{\mathrm{pos}}^{\textsc{st}}  \sqrt{\frac{\sigma_{\NHtwo}^2}{4\NHtwo^2}+\frac{\sigma_{\scriptscriptstyle W_{fil}}^2}{4 W_{fil}^2}+\frac{\sigma_{\scriptscriptstyle \Delta V_{NT}}^2}{\Delta V_{NT}^2}+\frac{\sigma_{\scriptscriptstyle \delta\theta}^2}{4\delta\theta^2}} .
\end{eqnarray}
In the above calculation, the standard error on the mean for column density was
calculated by the common expression :
$\sigma_{mean}$=Standard\,Deviation/$\sqrt{\mathrm{Data\,Points\,or\,Pixels}}$.
For the error on the filament width, we assume a normal distribution of the
filament widths, and use the interquartile range of 0.07\,pc provided by
\citet{Arzoumanian_FilWidth_2019AA} to obtain
$\sigma_{\scriptscriptstyle W_{fil}} = 0.07/1.349 \approx 0.05$\,pc.
According to \citet{Arzoumanian_2013AA}, the error on velocity dispersion
measurements are of the order of $\sim$0.05\,km\,s$^{-1}$. Thus the error on
velocity \textsc{fwhm} was taken to be
$\sigma_{\scriptscriptstyle \Delta V_{NT}}$=0.05$\times\sqrt{\mathrm{(8 ln2)}}$
$\approx$0.12\,\kmps.
For $\sigma_{\scriptscriptstyle \delta\theta}$, we utilise the standard error
on the circular mean (calculated as per appendix \ref{appendix_Errors_CircMean})
as an estimate for our calculation.

%%%%%%%%%%%%%%%%%%%%%%%%%%%%%%%%%%%%%%%%%%%%%%%%%%%%%%%%%%%%%%%%%%%%%%%%%%%%%%%%
\clearpage
\bibliography{bibliography.bib}{}
\bibliographystyle{aasjournal}
%%%%%%%%%%%%%%%%%%%%%%%%%%%%%%%%%%%%%%%%%%%%%%%%%%%%%%%%%%%%%%%%%%%%%%%%%%%%%%%%
\end{document}